\documentclass[letterpaper,12pt]{JHEP3}
\usepackage{amsmath,amssymb}
\usepackage{latexsym}
\usepackage{epsfig}
\usepackage{cite}
\usepackage[english]{babel}

\newcommand{\eq}{\begin{equation}}
\newcommand{\feq}{\end{equation}}
\newcommand{\eqn}{\begin{eqnarray}}
\newcommand{\feqn}{\end{eqnarray}}
\newcommand{\arr}{\begin{eqnarray*}}
\newcommand{\farr}{\end{eqnarray*}}

\newcommand{\DD}{{\cal D}}

\font\mybb=msbm10 at 12pt
\def\bb#1{\hbox{\mybb#1}}

\def\bI {\bb{I}}
\def\bR {\bb{R}}

\def\bC {\bb{C}}

\title{All timelike supersymmetric solutions of ${\cal N}=2$, $D=4$ gauged supergravity
coupled to abelian vector multiplets}

\author{Sergio L.~Cacciatori,$^{ac}$ Dietmar Klemm,$^{bc}$ Diego S.~Mansi$^d$ and
Emanuele Zorzan$^{bc}$ \\
$^a$ Dipartimento di Scienze Fisiche e Matematiche, \\
\hspace*{0.15cm} Universit\`a dell'Insubria, \\
\hspace*{0.15cm} Via Valleggio 11, I-22100 Como. \\
$^b$ Dipartimento di Fisica dell'Universit\`a di Milano, \\
\hspace*{0.15cm} Via Celoria 16, I-20133 Milano. \\
$^c$ INFN, Sezione di Milano, Via Celoria 16, I-20133 Milano. \\
$^d$ Department of Physics, University of Crete, \\
\hspace*{0.15cm} 71003 Heraklion, Greece. \\
}
\preprint{IFUM-917-FT}

\abstract{The timelike supersymmetric solutions of ${\cal N}=2$, $D=4$ gauged
supergravity coupled to an arbitrary number of abelian vector multiplets are
classified using spinorial geometry techniques. We show that the generalized holonomy
group for vacua preserving N supersymmetries is
GL$(\frac{8-N}2,\bC)$ $\ltimes\frac N2\bC^{\frac{8-N}2}$ $\subseteq$ GL$(8,\bR)$, where
$N=0,2,4,6,8$. The spacetime turns out to be a fibration over a three-dimensional base
manifold with U(1) holonomy and nontrivial torsion. Our results can be used to construct
new supersymmetric AdS black holes with nontrivial scalar fields turned on.}


\begin{document}

\section{Introduction}
\label{intro}

Supersymmetric solutions to supergravity theories have played, and continue to
play, an important role in string- and M-theory developments. This makes it
desirable to obtain a complete classification of BPS solutions to various
supergravities in diverse dimensions. Progress in this direction has been made in
the last years using the mathematical concept of G-structures \cite{Gauntlett:2002sc}.
The basic strategy is to assume the existence of at least one Killing spinor
$\epsilon$ obeying ${\cal D}_{\mu}\epsilon=0$, and to
construct differential forms as bilinears from this spinor. These forms, which define
a preferred G-structure, obey several algebraic and differential equations that can be
used to deduce the metric and the other bosonic supergravity fields.
Using this framework, a number of complete
classifications \cite{Gauntlett:2002nw, Gutowski:2003rg, Meessen:2006tu} and many
partial results (see e.g.~\cite{Gauntlett:2002fz, Gauntlett:2003fk, Caldarelli:2003pb,
Caldarelli:2003wh, Gauntlett:2003wb, Cariglia:2004kk, Cacciatori:2004rt, Cariglia:2004qi,
Gutowski:2005id, Bellorin:2005zc, Huebscher:2006mr, Bellorin:2006yr, Bellorin:2007yp}
for an incomplete list) have been obtained. By complete we mean that the most general
solutions for all possible fractions of supersymmetry have been obtained,
while for partial classifications this is only available for some
fractions. Note that the complete classifications mentioned above involve
theories with eight supercharges and holonomy $H=$ SL(2,$\mathbb{H}$) of the
supercurvature $R_{\mu\nu} = {\cal D}_{[\mu} {\cal D}_{\nu ]}$, and allow for either
half- or maximally supersymmetric solutions.

An approach which exploits the linearity of the Killing spinors has
been proposed \cite{Gillard:2004xq} under the name of spinorial
geometry. Its basic ingredients are an explicit oscillator basis for
the spinors in terms of forms and the use of the gauge symmetry to
transform them to a preferred representative of their orbit.
In this way one can construct a linear system for the background
fields from any (set of) Killing spinor(s) \cite{Gran:2005wu}. This
method has proven fruitful in e.g.~the challenging case of IIB
supergravity \cite{Gran:2005wn, Gran:2005kg, Gran:2005ct}. In
addition, it has been adjusted to impose 'near-maximal'
supersymmetry and thus has been used to rule out certain large fractions of
supersymmetry \cite{Gran:2006ec, Grover:2006ps, Grover:2006wy,
Gran:2006cn, Gran:2007eu}. Finally, a complete classification for type I supergravity
in ten dimensions has been obtained in \cite{Gran:2007fu}, and all half-supersymmetric
backgrounds of ${\cal N}=2$, $D=5$ gauged supergravity coupled to abelian
vector multiplets were determined in \cite{Gutowski:2007ai, Grover:2008ih}.

In the present paper we would like to address the classification of
supersymmetric solutions in four-dimensional ${\cal N}=2$ matter-coupled U(1)-gauged
supergravity, generalizing thus the simpler cases of ${\cal N}=1$, considered recently
in \cite{Ortin:2008wj,Gran:2008vx}, and minimal ${\cal N}=2$, where a full
classification is available both in the ungauged \cite{Tod:1983pm}
and gauged theories \cite{Cacciatori:2007vn}. We shall thereby focus on
the class where the Killing vector constructed from the Killing spinor is
timelike, deferring the lightlike case to a forthcoming publication.
Moreover, only coupling to abelian vector multiplets and gauging of a U(1) subgroup
of the SU(2) R-symmetry will be considered, while the inclusion of hypermultiplets and
nonabelian vectors, as well as a general gauging, are left for future work \cite{chkmmovz}.

The outline of this paper is as follows. In section \ref{sugra}, we briefly review
${\cal N}=2$ supergravity in four dimensions and its matter couplings. In \ref{orbits}
we discuss the orbits of Killing spinors and analyze the holonomy of the
supercovariant connection. In section \ref{timelike} we determine the conditions
coming from a single timelike Killing spinor, and obtain all supersymmetric
solutions in this class. Finally, in section \ref{finalrem} we present our conclusions
and outlook. Appendices \ref{conv} and \ref{spin-forms} contain
our notation and conventions for spinors, while in appendix \ref{BPS-eom} we
show that the Killing spinor equations, together with the Maxwell equations and the
Bianchi identities, imply the equations of motion in the timelike case.
Finally, in appendix \ref{hol-base} we discuss the reduced holonomy of the
three-dimensional manifold over which the spacetime is fibered.

\section{Matter-coupled ${\cal N}=2$, $D=4$ gauged supergravity}
\label{sugra}

In this section we shall give a short summary
of the main ingredients of ${\cal N}=2$, $D=4$ gauged supergravity coupled
to vector- and hypermultiplets \cite{Andrianopoli:1996cm}. Throughout this paper,
we will use the notations and conventions of \cite{Vambroes}, to which we refer
for more details.

Apart from the vierbein $e^a_{\mu}$ and the chiral gravitinos $\psi^i_{\mu}$,
$i=1,2$, the field content includes $n_H$ hypermultiplets and $n_V$ vector multiplets
enumerated by $I=0,\ldots,n_V$. The latter contain the graviphoton and have
fundamental vectors $A^I_{\mu}$, with field strengths
\begin{displaymath}
F^I_{\mu\nu} = \partial_{\mu}A^I_{\nu} - \partial_{\nu}A^I_{\mu} + gA^K_{\nu}A^J_{\mu}
              {f_{JK}}^I\,.
\end{displaymath}
The fermions of the vector multiplets are denoted as $\lambda^{\alpha i}$ and the
complex scalars as $z^{\alpha}$ where $\alpha=1,\ldots,n_V$. These scalars parametrize
a special K\"ahler manifold, i.~e.~, an $n_V$-dimensional
Hodge-K\"ahler manifold that is the base of a symplectic bundle, with the
covariantly holomorphic sections
\begin{equation}
{\cal V} = \left(\begin{array}{c} X^I \\ F_I\end{array}\right)\,, \qquad
{\cal D}_{\bar\alpha}{\cal V} = \partial_{\bar\alpha}{\cal V}-\frac 12
(\partial_{\bar\alpha}{\cal K}){\cal V}=0\,, \label{sympl-vec}
\end{equation}
where ${\cal K}$ is the K\"ahler potential and ${\cal D}$ denotes the
K\"ahler-covariant derivative\footnote{For a generic field $\phi^{\alpha}$ that
transforms under a K\"ahler transformation ${\cal K}(z,\bar z)\to
{\cal K}(z,\bar z)+\Lambda(z)+\bar\Lambda(\bar z)$ as
$\phi^{\alpha}\to e^{-(p\Lambda+q\bar\Lambda)/2}\phi^{\alpha}$, one has
${\cal D}_{\alpha}\phi^{\beta}=\partial_{\alpha}\phi^{\beta}+
{\Gamma^{\beta}}_{\alpha\gamma}\phi^{\gamma}+\frac p2(\partial_{\alpha}{\cal K})
\phi^{\beta}$. ${\cal D}_{\bar\alpha}$ is defined in the same
way. $X^I$ transforms as $X^I\to e^{-(\Lambda -\bar\Lambda)/2}X^I$ and
thus has K\"ahler weights $(p,q)=(1,-1)$.}.
${\cal V}$ obeys the symplectic constraint
\begin{equation}
\langle {\cal V}\,,\bar{\cal V}\rangle = X^I\bar F_I-F_I\bar X^I=i\,.
\end{equation}
To solve this condition, one defines
\begin{equation}
{\cal V}=e^{{\cal K}(z,\bar z)/2}v(z)\,,
\end{equation}
where $v(z)$ is a holomorphic symplectic vector,
\begin{equation}
v(z) = \left(\begin{array}{c} Z^I(z) \\ \frac{\partial}{\partial Z^I}F(Z)
\end{array}\right)\,.
\end{equation}
F is a homogeneous function of degree two, called the prepotential,
whose existence is assumed to obtain the last expression.
This is not restrictive because it can be shown
that it is always possible to go in a gauge where the prepotential exists via a local symplectic
transformation \cite{Vambroes,Craps:1997gp}\footnote{This need not
be true for gauged supergravity, where symplectic covariance is broken
\cite{Andrianopoli:1996cm}. However, in our analysis we do not really use
that the $F_I$ can be obtained from a prepotential, so
our conclusions go through also without assuming that $F_I=\partial
F(X)/\partial X^I$ for some $F(X)$.
We would like to thank Patrick Meessen for discussions on this point.}.
The K\"ahler potential is then
\begin{equation}
e^{-{\cal K}(z,\bar z)} = -i\langle v\,,\bar v\rangle\,.
\end{equation}
The matrix ${\cal N}_{IJ}$ determining the coupling between the scalars $z^{\alpha}$
and the vectors $A^I_{\mu}$ is defined by the relations
\begin{equation}\label{defN}
F_I = {\cal N}_{IJ}X^J\,, \qquad {\cal D}_{\bar\alpha}\bar F_I = {\cal N}_{IJ}
{\cal D}_{\bar\alpha}\bar X^J\,.
\end{equation}
Given
\begin{equation}
U_{\alpha} \equiv {\cal D}_{\alpha}{\cal V} = \partial_{\alpha}{\cal V} + \frac 12
(\partial_{\alpha}{\cal K}){\cal V}\,,
\end{equation}
the following differential constraints hold:
\begin{eqnarray}
{\cal D}_{\alpha}U_{\beta} &=& C_{\alpha\beta\gamma}g^{\gamma\bar\delta}\bar U_{\bar\delta}\,,
\nonumber \\
{\cal D}_{\bar\beta}U_{\alpha} &=& g_{\alpha\bar\beta}{\cal V}\,, \nonumber \\
\langle U_{\alpha}\,, {\cal V}\rangle &=& 0\,.
\end{eqnarray}
Here, $C_{\alpha\beta\gamma}$ is a completely symmetric tensor which determines also
the curvature of the special K\"ahler manifold.

We now come to the hypermultiplets. These contain scalars $q^X$ and spinors
$\zeta^A$, where $X=1,\ldots,4n_H$ and $A=1,\ldots,2n_H$. The $4n_H$ hyperscalars
parametrize a quaternionic K\"ahler manifold, with vielbein $f^{iA}_X$ and inverse
$f^X_{iA}$ (i.~e.~the tangent space is labelled by indices $(iA)$). From these one
can construct the three complex structures
\begin{equation}
\vec J_X^{\;\;\;Y} = -if^{iA}_X\vec\sigma_i^{\;\;j} f^Y_{jA}\,,
\end{equation}
with the Pauli matrices $\vec\sigma_i^{\;\;j}$ (cf.~appendix \ref{conv}).
Furthermore, one defines SU$(2)$ connections $\vec\omega_X$ by requiring the
covariant constancy of the complex structures:
\begin{equation}
0 = {\mathfrak D}_X\vec J_Y^{\;\;\;Z}\equiv \partial_X\vec J_Y^{\;\;\;Z}-
{\Gamma^W}_{XY}\vec J_W^{\;\;\;Z}+{\Gamma^Z}_{XW}\vec J_Y^{\;\;\;W}+
2\,\vec\omega_X\times\vec J_Y^{\;\;\;Z}\,,
\end{equation}
where the Levi-Civita connection of the metric $g_{XY}$ is used. The curvature
of this SU$(2)$ connection is related to the complex structure by
\begin{equation}
{\vec R}_{XY} \equiv 2\,\partial_{\left[X\right.}\vec\omega_{\left.Y\right]} +
2\,\vec\omega_X\times\vec\omega_Y = -\frac 12\kappa^2{\vec J}_{XY}\,.
\end{equation}
Depending on whether $\kappa=0$ or $\kappa\neq 0$ the manifold is hyper-K\"ahler
or quaternionic K\"ahler respectively. In what follows, we take $\kappa=1$.

The bosonic action of ${\cal N}=2$, $D=4$ supergravity is
\begin{eqnarray}
e^{-1}{\cal L}_{\text{bos}} &=& \frac 1{16\pi G}R + \frac 14(\text{Im}\,{\cal N})_{IJ}
F^I_{\mu\nu}F^{J\mu\nu} - \frac 18(\text{Re}\,{\cal N})_{IJ}\,e^{-1}
\epsilon^{\mu\nu\rho\sigma}F^I_{\mu\nu}F^J_{\rho\sigma}\,, \nonumber \\
&& -g_{\alpha\bar\beta}{\cal D}_{\mu}z^{\alpha}{\cal D}^{\mu}\bar z^{\bar\beta} -
\frac 12 g_{XY}{\cal D}_{\mu}q^X{\cal D}^{\mu}q^Y - V\,, \nonumber \\
&& -\frac g6 C_{I,JK}e^{-1}\epsilon^{\mu\nu\rho\sigma}A^I_{\mu}A^J_{\nu}
(\partial_{\rho}A^K_{\sigma}-\frac 38 g{f_{LM}}^KA^L_{\rho}A^M_{\sigma})\,,
\label{action}
\end{eqnarray}
where $C_{I,JK}$ are real coefficients, symmetric in the last two indices,
with $Z^I Z^J Z^K C_{I,JK}=0$, and the covariant derivatives acting on the
scalars read
\begin{equation}
{\cal D}_{\mu}z^{\alpha} = \partial_{\mu}z^{\alpha} + g A^I_{\mu}k^{\alpha}_I(z)\,,
\qquad {\cal D}_{\mu}q^X = \partial_{\mu}q^X + g A^I_{\mu}k^X_I\,.
\end{equation}
Here $k^{\alpha}_I(z)$ and $k^X_I$ are Killing vectors of the special K\"ahler
and quaternionic K\"ahler manifolds respectively. The potential $V$
in \eqref{action} is the sum of three distinct contributions:
\begin{eqnarray}
V &=& g^2(V_1+V_2+V_3)\,, \nonumber \\
V_1 &=& g_{\alpha\bar\beta}k^{\alpha}_I k^{\bar\beta}_J\,e^{\cal K}\bar Z^I Z^J\,,
\nonumber \\
V_2 &=& 2\,g_{XY}k^X_I k^Y_Je^{\cal K}\bar Z^I Z^J\,, \nonumber \\
V_3 &=& 4(U^{IJ}-3\,e^{\cal K}\bar Z^I Z^J)\vec P_I\cdot\vec P_J\,, \label{scal-pot}
\end{eqnarray}
with
\begin{equation}
U^{IJ} \equiv g^{\alpha\bar\beta}e^{\cal K}{\cal D}_{\alpha}Z^I{\cal D}_{\bar\beta}
\bar Z^J = -\frac 12(\text{Im}\,{\cal N})^{-1|IJ}-e^{\cal K}\bar Z^I Z^J\,,
\end{equation}
and the triple moment maps $\vec P_I(q)$. The latter have to satisfy the
equivariance condition
\begin{equation}
\vec P_I\times\vec P_J + \frac 12 \vec J_{XY}k^X_Ik^Y_J - {f_{IJ}}^K
\vec P_K = 0\,, \label{equivariance}
\end{equation}
which is implied by the algebra of symmetries. The metric for the vectors
is given by
\begin{equation}
{\cal N}_{IJ}(z,\bar z) = \bar F_{IJ} + i\frac{N_{IN}N_{JK}Z^NZ^K}{N_{LM}Z^LZ^M}\,,
\qquad N_{IJ} \equiv 2\,\text{Im}\,F_{IJ}\,,
\end{equation}
where $F_{IJ}=\partial_I\partial_JF$, and $F$ denotes the prepotential.

Finally, the supersymmetry transformations of the fermions to bosons are
\begin{eqnarray}
\delta\psi^i_{\mu} &=& D_{\mu}(\omega)\epsilon^i - g\Gamma_{\mu}S^{ij}\epsilon_j
+ \frac 14\Gamma^{ab}F^{-I}_{ab}\epsilon^{ij}\Gamma_{\mu}\epsilon_j(\text{Im}\,
{\cal N})_{IJ}Z^Je^{{\cal K}/2}\,, \label{deltapsi} \\
D_{\mu}(\omega)\epsilon^i &=& (\partial_{\mu}+\frac 14\omega^{ab}_{\mu}\Gamma_{ab})
\epsilon^i + \frac i2 A_{\mu}\epsilon^i + \partial_{\mu}q^X{\omega_{X\,j}}^i\epsilon^j
+ g A_{\mu}^I{P_{I\,j}}^i\epsilon^j\,, \label{derivata} \\
\delta\lambda^{\alpha}_i &=& -\frac 12 e^{{\cal K}/2}g^{\alpha\bar\beta}
{\cal D}_{\bar\beta}\bar Z^I(\text{Im}\,{\cal N})_{IJ}F^{-J}_{\mu\nu}\Gamma^{\mu\nu}
\epsilon_{ij}\epsilon^j + \Gamma^{\mu}{\cal D}_{\mu}z^{\alpha}\epsilon_i
+ gN^{\alpha}_{ij}\epsilon^j\,, \nonumber \\
\delta\zeta^A &=& \frac i2 f^{Ai}_X\Gamma^{\mu}{\cal D}_{\mu}q^X\epsilon_i +
g{\cal N}^{iA}\epsilon_{ij}\epsilon^j\,, \nonumber
\end{eqnarray}
where we defined
\begin{eqnarray}
S^{ij} &\equiv& -P^{ij}_Ie^{{\cal K}/2}Z^I\,, \nonumber \\
N^{\alpha}_{ij} &\equiv& e^{{\cal K}/2}\left[\epsilon_{ij}k^{\alpha}_I\bar Z^I -
2P_{Iij}{\cal D}_{\bar\beta}\bar Z^Ig^{\alpha\bar\beta}\right]\,, \qquad
{\cal N}^{iA} \equiv -if^{iA}_Xk^X_Ie^{{\cal K}/2}\bar Z^I\,. \nonumber
\end{eqnarray}
In \eqref{derivata}, $A_{\mu}$ is the gauge field of the K\"ahler U$(1)$,
\begin{equation}
A_{\mu} = -\frac i2(\partial_{\alpha}{\cal K}\partial_{\mu}z^{\alpha} -
\partial_{\bar\alpha}{\cal K}\partial_{\mu}\bar z^{\bar\alpha}) - gA^I_{\mu}P^0_I\,,
\label{KaehlerU1}
\end{equation}
with the moment map function
\begin{equation}
P^0_I = \langle T_I {\cal V}\,,\bar{\cal V}\rangle\,, \label{P0I}
\end{equation}
and
\begin{equation}
T_I{\cal V} \equiv\left(\begin{array}{cc} -{f_{IJ}}^K & 0 \\ C_{I,KJ} & {f_{IK}}^J
\end{array}\right)\left(\begin{array}{c} X^J \\ F_J\end{array}\right)\,.
\label{TIV}
\end{equation}
The major part of this paper will deal with the case of vector multiplets only,
i.~e.~, $n_H=0$. Then there are still two possible solutions of \eqref{equivariance}
for the moment maps $\vec P_I$, which are called SU(2) and U(1) Fayet-Iliopoulos
(FI) terms respectively \cite{Vambroes}. Here we are interested in the latter.
In this case
\begin{equation}
\vec P_I = \vec e\,\xi_I\,, \label{FI}
\end{equation}
where $\vec e$ is an arbitrary vector in SU(2) space and $\xi_I$ are constants
for the $I$ corresponding to U(1) factors in the gauge group.
If, moreover, we assume ${f_{IJ}}^K=0$ (abelian gauge group), and $k^{\alpha}_I=0$
(no gauging of special K\"ahler isometries), then only the $V_3$ part survives
in the scalar potential \eqref{scal-pot}, and one can also choose $C_{I,JK}=0$.
Note that this case corresponds to a gauging of a U(1) subgroup of the SU(2)
R-symmetry, with gauge field $\xi_IA^I_{\mu}$.

\section{$G$-invariant Killing spinors in 4D}

\subsection{Orbits of spinors under the gauge group}
\label{orbits}

A Killing spinor\footnote{Our conventions for spinors and their description in terms
of forms can be found in appendix \ref{spin-forms}.} can be viewed as an SU(2) doublet
$(\epsilon^1, \epsilon^2)$, where an upper index means that a spinor has positive
chirality. $\epsilon^i$ is related to the negative chirality spinor $\epsilon_i$ by
charge conjugation, $\epsilon_i^C = \epsilon^i$, with
\begin{equation}
\epsilon_i^C = \Gamma_0 C^{-1}\epsilon_i^{\ast}\,.
\end{equation}
Here $C$ is the charge conjugation matrix defined in appendix \ref{spin-forms}.
As $\epsilon^1$ has positive chirality, we can write $\epsilon^1 = c1 + de_{12}$
for some complex functions $c,d$. Notice that $c1 + de_{12}$ is in the same orbit as 1
under Spin(3,1), which can be seen from
\begin{displaymath}
e^{\gamma\Gamma_{13}}e^{\psi\Gamma_{12}}e^{\delta\Gamma_{13}}e^{h\Gamma_{02}}\,1
= e^{i(\delta+\gamma)}e^h\cos\psi\,1 + e^{i(\delta-\gamma)}e^h\sin\psi\,e_{12}\,.
\end{displaymath}
This means that we can set $c=1$, $d=0$ without loss of generality. In order to
determine the stability subgroup of $\epsilon^1$, one has to solve the
infinitesimal equation
\begin{equation}
\alpha^{cd}\Gamma_{cd}1 = 0\,, \label{stab}
\end{equation}
which implies $\alpha^{02} = \alpha^{13} = 0$, $\alpha^{01} = -\alpha^{12}$,
$\alpha^{03} = \alpha^{23}$. The stability subgroup of 1 is thus generated by
\begin{equation}
X = \Gamma_{01} - \Gamma_{12}\,, \qquad Y = \Gamma_{03} + \Gamma_{23}\,. \label{XY}
\end{equation}
One easily verifies that $X^2 = Y^2 = XY = 0$, and thus $\exp(\mu X
+ \nu Y) = 1 + \mu X + \nu Y$, so that $X,Y$ generate $\bR^2$.

Having fixed $\epsilon^1 = 1$, also $\epsilon_1$ is determined by
$\epsilon_1 = \epsilon^{1C} = e_1$. A negative chirality spinor independent of
$\epsilon^1$ is $\epsilon_2$, which can be written as a linear combination of
odd forms, $\epsilon_2 = ae_1 + be_2$, where $a$ and $b$ are again complex valued
functions. We can now act with the stability subgroup of $\epsilon^1$ to bring
$\epsilon_2$ to a special form:
\begin{displaymath}
(1 + \mu X + \nu Y)(ae_1 + be_2) = be_2 + [a - 2b(\mu + i\nu)]e_1\,.
\end{displaymath}
In the case $b=0$ this spinor is invariant, so the representative is
$\epsilon^1=1$, $\epsilon_2=ae_1$ (so that $\epsilon^2 = \bar a 1$), with isotropy
group $\bR^2$. If $b \neq 0$, one can bring the spinor to the form $be_2$
(which implies $\epsilon^2 = -\bar b e_{12}$), with isotropy group $\bI$.
The representatives\footnote{Note the difference in form compared to the Killing spinors
of the corresponding theories in five and six dimensions: in six dimensions these can be
chosen constant \cite{Gutowski:2003rg} while in five dimensions they are constant up to
an overall function \cite{Grover:2006ps}. In four dimensions such a choice is generically
not possible.} together with the stability subgroups are
summarized in table \ref{tab:orbits}.
Given a Killing spinor $\epsilon^i$, one can construct the bilinear
\begin{equation}
V_A = A(\epsilon^i,\Gamma_A\epsilon_i)\,,
\end{equation}
with the Majorana inner product $A$ defined in \eqref{Majorana}, and the sum
over $i$ is understood. For $\epsilon_2 = ae_1$, $V_A$ is lightlike, whereas for
$\epsilon_2 = be_2$ it is timelike, see table~\ref{tab:orbits}. The existence of a
globally defined Killing spinor $\epsilon^i$, with isotropy group $G \in$ Spin(3,1),
gives rise to a $G$-structure. This means that we have an
$\bR^2$-structure in the null case and an identity structure in the
timelike case.

In U(1) gauged supergravity, the local Spin(3,1) invariance is
actually enhanced to Spin(3,1) $\times$ U(1). For U(1) Fayet-Iliopoulos terms,
the moment maps satisfy \eqref{FI}, where we can choose $e^x = \delta^x_3$
without loss of generality. Then, under a gauge transformation
\begin{equation}
A^I_{\mu} \to A^I_{\mu} + \partial_{\mu}\alpha^I\,, \label{U(1)transf}
\end{equation}
the Killing spinor $\epsilon^i$ transforms as
\begin{equation}
\epsilon^1 \to e^{-ig\xi_I\alpha^I}\epsilon^1\,, \qquad
\epsilon^2 \to e^{ig\xi_I\alpha^I}\epsilon^2\,, \label{transf_eps}
\end{equation}
which can be easily seen from the supercovariant derivative
(cf.~eq.~\eqref{derivata}). Note that
$\epsilon^1$ and $\epsilon^2$ have opposite charges under the U(1).
In order to obtain the stability subgroup, one determines the Lorentz
transformations that leave the spinors $\epsilon^1$ and $\epsilon^2$ invariant
up to a arbitrary phase factors $e^{i\psi}$ and $e^{-i\psi}$ respectively, which can
then be gauged away using the additional U(1) symmetry.
If $\epsilon_2=0$, one gets in this way an
isotropy group generated by $X, Y$ and $\Gamma_{13}$ obeying
\begin{displaymath}
[\Gamma_{13}, X] = -2Y\,, \qquad [\Gamma_{13}, Y] = 2X\,, \qquad [X, Y] = 0\,,
\end{displaymath}
i.~e.~$G\cong$ U(1)$\ltimes \bR^2$. For $\epsilon_2 = ae_1$ with $a \neq 0$,
the stability subgroup $\bR^2$ is not enhanced, whereas the $\bI$ of
the representative $(\epsilon^1,\epsilon_2) = (1,be_2)$ is promoted to U(1)
generated by $\Gamma_{13} = i\Gamma_{\bar\bullet\bullet}$. The Lorentz
transformation matrix $a_{AB}$ corresponding to $\Lambda =
\exp(i\psi\Gamma_{\bar\bullet\bullet}) \in$ U(1), with
$\Lambda\Gamma_B\Lambda^{-1} = {a^A}_B\Gamma_A$, has nonvanishing
components
\begin{equation}
a_{+-} = a_{-+} = 1\,, \qquad a_{\bullet\bar\bullet} = e^{2i\psi}\,, \qquad
a_{\bar\bullet\bullet} = e^{-2i\psi}\,. \label{residualU1}
\end{equation}
Finally, notice that in U(1) gauged supergravity one can choose the
function $a$ in $\epsilon_2=ae_1$ real and positive: Write $a =
R\exp(2i\delta)$, use
\begin{displaymath}
e^{\delta\Gamma_{13}}1 = e^{i\delta}1\,, \qquad
e^{\delta\Gamma_{13}}ae_1 = e^{-i\delta}ae_1 = e^{i\delta}Re_1\,,
\end{displaymath}
and gauge away the phase factor $\exp(i\delta)$ using the
electromagnetic U(1).

\begin{table}[ht]
\begin{center}
\begin{tabular}{||c||c|c||c||}
\hline $(\epsilon^1,\epsilon_2)$ & $G\subset$ Spin(3,1) & $G\subset$ Spin(3,1)
$\times$ U(1) &
$V_A E^A = A(\epsilon^i,\Gamma_A\epsilon_i)E^A$ \\
\hline\hline
$(1,0)$ & $\bR^2$ & U(1)$\ltimes \bR^2$$\,\,\,\,$ & $-\sqrt 2 E^-$ \\
\hline
$(1,ae_1)$ & $\bR^2$ & $\bR^2 \,\,\,\, (a \in \bR)$ & $-\sqrt 2(1+a^2)E^-$ \\
\hline
$(1,be_2)$ & $\bI$ & U(1)$\,\,\,\,$ &  $\sqrt 2(|b|^2E^+-E^-)$ \\
\hline
\end{tabular}
\end{center}
\caption{The representatives $(\epsilon^1,\epsilon_2)$ of the orbits of Weyl
spinors and their stability subgroups $G$ under the gauge groups
Spin(3,1) and Spin(3,1) $\times$ U(1) in the ungauged and U(1)-gauged
theories, respectively. The number of orbits is the same in both
theories, the only difference lies in the stability subgroups and
the fact that $a$ is real in the gauged theory. In the last column
we give the vectors constructed from the spinors.}\label{tab:orbits}
\end{table}

Note that in the gauged theory the presence of $G$-invariant Killing
spinors will in general not lead to a $G$-structure on the manifold
but to stronger conditions. The structure group is in fact reduced
to the intersection of $G$ with Spin(3,1), and hence is equal to
the stability subgroup in the ungauged theory.

The representatives, stability subgroups and vectors constructed from
the Killing spinors are summarized in table \ref{tab:orbits} both for
the ungauged and the U(1)-gauged cases.



\subsection{Generalized holonomy}

The variation of the chiral gravitini under supersymmetry transformations is given
by \eqref{deltapsi}. This can be rewritten in terms of Majorana spinors
$\psi^{\underline i}_{\mu}=\psi^i_{\mu}+\psi_{i\mu}$ and
$\epsilon^{\underline i}=\epsilon^i+\epsilon_i$, where $\psi_{i\mu}$ and
$\epsilon_i$ denote the charge conjugate of $\psi^i_{\mu}$ and $\epsilon^i$
respectively. One has then
\begin{eqnarray}
\delta\psi^{\underline i}_{\mu} &=& {\hat{\cal D}}_{\mu}\epsilon^{\underline i} =
(\partial_{\mu}+\frac 14\omega^{ab}_{\mu}\Gamma_{ab})\epsilon^{\underline i} +
\frac i2 A_{\mu}\Gamma_5\epsilon^{\underline i} +
\partial_{\mu}q^X\left[\text{Re}\,{\omega_{X\,j}}^i +
i\Gamma_5\text{Im}\,{\omega_{X\,j}}^i\right]\epsilon^{\underline j}\nonumber \\
&& + g A^I_{\mu}\left[\text{Re}{P_{I\,j}}^i + i\Gamma_5\text{Im}{P_{I\,j}}^i
\right]\epsilon^{\underline j} + g\Gamma_{\mu}e^{{\cal K}/2}\left[\text{Re}(P^{ij}_I
Z^I) - i\Gamma_5\text{Im}(P^{ij}_IZ^I)\right]\epsilon^{\underline j}\nonumber \\
&& + \frac 14\Gamma\cdot\left[\text{Re}(F^{-I}Z^J) + i\Gamma_5\text{Im}(F^{-I}
Z^J)\right]\epsilon^{ij}\Gamma_{\mu}\epsilon^{\underline j}(\text{Im}\,{\cal N})_{IJ}
e^{{\cal K}/2}\,.
\end{eqnarray}
From this it is evident that the holonomy of the supercovariant derivative
${\hat{\cal D}}_{\mu}$ is contained in GL$(8,\bR)$, so that in principle one can
have vacua that preserve any number $N$ of supersymmetries with $N=0,1,\ldots 8$.
In the case without hypermultiplets, and for U(1) FI terms with
${\vec P}_I=\vec e\,\xi_I$ and $e^x = \delta^x_3$, it is instructive to rewrite
everything using complex (Dirac) spinors $\psi_{\mu}=\psi^1_{\mu}+\psi_{2\mu}$,
$\epsilon=\epsilon^1+\epsilon_2$\footnote{Note that one can reconstuct $\psi^1_{\mu}$
and $\psi_{2\mu}$ from $\psi_{\mu}$ by projecting on the two chiralities.}.
This yields
\begin{eqnarray}
\delta\psi_{\mu} &=& (\partial_{\mu}+\frac 14\omega^{ab}_{\mu}\Gamma_{ab})\epsilon
+ \frac i2 A_{\mu}\Gamma_5\epsilon + ig\xi_IA^I_{\mu}\epsilon
+g\Gamma_{\mu}\xi_I\left[\text{Im}X^I + i\Gamma_5\text{Re}X^I\right]\epsilon\nonumber \\
&& +\frac i4\Gamma\cdot\left[\text{Im}(F^{-I}X^J) - i\Gamma_5\text{Re}(F^{-I}X^J)
\right](\text{Im}\,{\cal N})_{IJ}\Gamma_{\mu}\epsilon
\end{eqnarray}
as well as (introducing $\lambda^{\alpha}=\lambda^{\alpha}_2+{\lambda^{\alpha}_1}^C$)
\begin{eqnarray}
&&\delta\lambda^{\alpha} = \frac i2 e^{{\cal K}/2}(\text{Im}\,{\cal N})_{IJ}
\Gamma\cdot\left[\text{Im}(F^{-J}{\cal D}_{\bar\beta}{\bar Z}^Ig^{\alpha\bar\beta})
- i\Gamma_5\text{Re}(F^{-J}{\cal D}_{\bar\beta}{\bar Z}^Ig^{\alpha\bar\beta})\right]
\epsilon\nonumber \\
&& + \Gamma^{\mu}\partial_{\mu}\left[\text{Re}z^{\alpha} -
i\Gamma_5\text{Im}z^{\alpha}\right]\epsilon + 2g e^{{\cal K}/2}\xi_I
\left[\text{Im}({\cal D}_{\bar\beta}{\bar Z}^Ig^{\alpha\bar\beta}) - i\Gamma_5
\text{Re}({\cal D}_{\bar\beta}{\bar Z}^Ig^{\alpha\bar\beta})\right]\epsilon\,. \nonumber
\end{eqnarray}
We see that in this case the complex conjugate spinor $\epsilon^{\ast}$ does not
appear in the variation of the fermions, so that the supercovariant derivative
has smaller holonomy GL$(4,\bC)$ $\subseteq$ GL$(8,\bR)$, and the number of preserved
supercharges is necessarily even, $N=0,2,4,6,8$. The generalized holonomy group
for vacua preserving $N$ supersymmetries is then
GL$(\frac{8-N}2,\bC)$ $\ltimes\frac N2\bC^{\frac{8-N}2}$, like in minimal gauged
supergravity \cite{Batrachenko:2004su,Cacciatori:2007vn}. To see this, assume that
there exists a Killing spinor $\epsilon_1$\footnote{The index of $\epsilon_1$ here
should not be confused with an SU(2) index for chiral spinors.}. By a local
GL$(4,\bC)$ transformation, $\epsilon_1$ can be
brought to the form $\epsilon_1 = (1,0,0,0)^T$. This is annihilated by matrices of
the form
\begin{displaymath}
\cal A = \left(\begin{array}{cc} 0 & {\underline a}^T \\
         \underline 0 & A \end{array}\right)\,,
\end{displaymath}
that generate the affine group A$(3,\bC) \cong$ GL$(3,\bC) \ltimes \bC^3$.
Now impose a second Killing spinor $\epsilon_2 =
(\epsilon_2^0,{\underline\epsilon}_2)^T$. Acting with the stability subgroup
of $\epsilon_1$ yields
\begin{displaymath}
e^{\cal A}\epsilon_2 = \left(\begin{array}{c} \epsilon_2^0 + {\underline b}^T
{\underline\epsilon}_2 \\ e^A{\underline\epsilon}_2\end{array}\right)\,,
\qquad\mathrm{where}\quad {\underline b}^T = {\underline a}^T A^{-1}(e^A - 1)\,.
\end{displaymath}
We can choose $A \in$ gl$(3,\bC)$ such that $e^A{\underline\epsilon}_2 = (1,0,0)^T$,
and $\underline b$ such that $\epsilon_2^0 + {\underline b}^T{\underline\epsilon}_2
= 0$. This means that the stability subgroup of $\epsilon_1$ can be used to bring
$\epsilon_2$ to the form $\epsilon_2 = (0,1,0,0)$. The subgroup of A$(3,\bC)$
that stabilizes also $\epsilon_2$ consists of the matrices
\begin{displaymath}
\left(\begin{array}{cccc} 1 & 0 & b_2 & b_3 \\ 0 & 1 & B_{12} & B_{13} \\
0 & 0 & B_{22} & B_{23} \\ 0 & 0 & B_{32} & B_{33} \end{array}\right)
\quad \in {\mathrm{GL}}(2,\bC) \ltimes 2\bC^2\,.
\end{displaymath}
Finally, imposing a third Killing spinor yields GL$(1,\bC) \ltimes 3\bC$
as maximal generalized holonomy group, which is however not realized in
${\cal N}=2$, $D=4$ minimal gauged
supergravity \cite{Cacciatori:2004rt,Grover:2006wy}\footnote{3/4 supersymmetric
solutions of minimal gauged supergravity are necessarily quotients of AdS$_4$,
which have been constructed in \cite{FigueroaO'Farrill:2007kb}.}.
It would be interesting to see whether genuine preons (i.e., 3/4 supersymmetric
backgrounds that are not locally AdS) exist in matter-coupled
supergravity.

\section{Timelike representative $(\epsilon^1,\epsilon_2)=(1,be_2)$}
\label{timelike}

In this section we will analyze the conditions coming from a single
timelike Killing spinor, and determine all supersymmetric
solutions in this class. We shall first keep things general, i.~e.~,
including hypermultiplets and a general gauging, and write down the
linear system following from the Killing spinor equations. This system will
then be solved for the case of U(1) Fayet-Iliopoulos terms and without hypers,
while the solution in the general case will be left for a future
publication \cite{chkmmovz}.

\subsection{Conditions from the Killing spinor equations}

From the vanishing of the hyperini variation one obtains
\begin{eqnarray}
\frac i{\sqrt 2}f^{A1}_X {\cal D}_{\bullet}q^X + \frac{ib}{\sqrt 2}f^{A2}_X {\cal D}_-q^X
- g\,{\cal N}^{2A} &=& 0\,, \\
-\frac i{\sqrt 2}f^{A1}_X {\cal D}_+q^X + \frac{ib}{\sqrt 2}f^{A2}_X {\cal D}_{\bar\bullet}
q^X - g\bar b\,{\cal N}^{1A} &=& 0\,,
\end{eqnarray}
whereas the gaugino variation yields
\begin{equation}
\bar b e^{{\cal K}/2}g^{\alpha\bar\beta}{\cal D}_{\bar\beta}{\bar Z}^I
(\text{Im}\,{\cal N})_{IJ}(F^{-J\bullet\bar\bullet}-F^{-J+-}) - \sqrt 2 {\cal D}_+
z^{\alpha} - g\bar b N^{\alpha}_{12} = 0\,, \label{G1}
\end{equation}
\begin{equation}
2\bar b e^{{\cal K}/2} g^{\alpha\bar\beta}{\cal D}_{\bar\beta}{\bar Z}^I
(\text{Im}\,{\cal N})_{IJ} F^{-J+\bar\bullet} + \sqrt 2 {\cal D}_{\bullet}
z^{\alpha} + g N^{\alpha}_{11} = 0\,, \label{G2}
\end{equation}
\begin{equation}
e^{{\cal K}/2}g^{\alpha\bar\beta}{\cal D}_{\bar\beta}{\bar Z}^I
(\text{Im}\,{\cal N})_{IJ}(F^{-J+-}-F^{-J\bullet\bar\bullet}) + b\sqrt 2 {\cal D}_-
z^{\alpha} + g N^{\alpha}_{21} = 0\,, \label{G3}
\end{equation}
\begin{equation}
-2e^{{\cal K}/2} g^{\alpha\bar\beta}{\cal D}_{\bar\beta}{\bar Z}^I
(\text{Im}\,{\cal N})_{IJ} F^{-J-\bullet} + b\sqrt 2 {\cal D}_{\bar\bullet}
z^{\alpha} - g\bar b N^{\alpha}_{22} = 0\,. \label{G4}
\end{equation}
Finally, from the gravitini we get
\begin{eqnarray}
\frac 12(\omega^{+-}_+ - \omega^{\bullet\bar\bullet}_+) + \frac i2 A_+ + \partial_+
q^X{\omega_{X\,1}}^1 + g A^I_+ {P_{I\,1}}^1 && \nonumber \\
 - \sqrt 2 gb S^{12} + \frac b{\sqrt2}(F^{-I+-}-F^{-I\bullet\bar\bullet})
(\text{Im}\,{\cal N})_{IJ}Z^J e^{{\cal K}/2} &=& 0\,, \label{grav+1}
\end{eqnarray}
\begin{equation}
-\omega^{-\bullet}_+ - \bar b\partial_+q^X{\omega_{X\,2}}^1 - g\bar b A^I_+
{P_{I\,2}}^1 - \sqrt 2 b F^{-I-\bullet}(\text{Im}\,{\cal N})_{IJ}Z^J e^{{\cal K}/2} = 0\,,
\end{equation}
\begin{equation}
-\bar b\,\omega^{+\bar\bullet}_+ + \partial_+q^X{\omega_{X\,1}}^2 + g A^I_+
{P_{I\,1}}^2 - \sqrt 2 g b S^{22} = 0\,, \label{grav+3}
\end{equation}
\begin{equation}
-\partial_+\bar b - \frac{\bar b}2(\omega^{\bullet\bar\bullet}_+ - \omega^{+-}_+)
-\frac{i\bar b}2 A_+ - \bar b\partial_+ q^X{\omega_{X\,2}}^2 - g\bar b A^I_+
{P_{I\,2}}^2 = 0\,, \label{grav+4}
\end{equation}

\begin{equation}
\frac 12(\omega^{+-}_- - \omega^{\bullet\bar\bullet}_-)
+\frac i2 A_- + \partial_- q^X{\omega_{X\,1}}^1 + g A^I_- {P_{I\,1}}^1 = 0\,,
\label{grav-1}
\end{equation}
\begin{equation}
-\omega^{-\bullet}_- - \bar b\partial_-q^X{\omega_{X\,2}}^1 - \bar b g A^I_-
{P_{I\,2}}^1 + \sqrt 2 g S^{11} = 0\,, \label{grav-2}
\end{equation}
\begin{eqnarray}
-\partial_-\bar b - \frac{\bar b}2(\omega^{\bullet\bar\bullet}_- - \omega^{+-}_-)
- \frac{i\bar b}2 A_- - \bar b\partial_-q^X{\omega_{X\,2}}^2 - \bar b g A^I_-
{P_{I\,2}}^2 && \nonumber \\
 + \sqrt 2 g S^{21} + \frac 1{\sqrt2}(F^{-I\bullet\bar\bullet}-F^{-I+-})
(\text{Im}\,{\cal N})_{IJ}Z^J e^{{\cal K}/2} &=& 0\,, \label{grav-3}
\end{eqnarray}
\begin{equation}
-\bar b\,\omega^{+\bar\bullet}_- + \partial_-q^X{\omega_{X\,1}}^2 + g A^I_-
{P_{I\,1}}^2 + \sqrt 2 F^{-I+\bar\bullet}(\text{Im}\,{\cal N})_{IJ}Z^J
e^{{\cal K}/2} = 0\,,
\end{equation}

\begin{equation}
\frac 12(\omega^{+-}_{\bullet} - \omega^{\bullet\bar\bullet}_{\bullet}) +
\frac i2 A_{\bullet} + \partial_{\bullet}q^X{\omega_{X\,1}}^1 + g A^I_{\bullet}
{P_{I\,1}}^1 + \sqrt2 b F^{-I+\bar\bullet}(\text{Im}\,{\cal N})_{IJ}Z^J
e^{{\cal K}/2} = 0\,,
\end{equation}
\begin{eqnarray}
-\omega^{-\bullet}_{\bullet} - \bar b\partial_{\bullet}q^X{\omega_{X\,2}}^1 - \bar b
g A^I_{\bullet}{P_{I\,2}}^1 - \sqrt 2 gb S^{12} && \nonumber \\
+\frac b{\sqrt2}(F^{-I\bullet\bar\bullet}-F^{-I+-})(\text{Im}\,{\cal N})_{IJ}Z^J
e^{{\cal K}/2} &=& 0\,,
\end{eqnarray}
\begin{equation}
-\bar b\,\omega^{+\bar\bullet}_{\bullet} + \partial_{\bullet}q^X{\omega_{X\,1}}^2
+ g A^I_{\bullet}{P_{I\,1}}^2 = 0\,, \label{gravbullet3}
\end{equation}
\begin{equation}
-\partial_{\bullet}\bar b - \frac{\bar b}2(\omega^{\bullet\bar\bullet}_{\bullet}
- \omega^{+-}_{\bullet}) - \frac{i\bar b}2 A_{\bullet} - \bar b\partial_{\bullet}
q^X{\omega_{X\,2}}^2 - g\bar b A^I_{\bullet}{P_{I\,2}}^2 -
\sqrt 2 gb S^{22} = 0\,,
\end{equation}

\begin{equation}
\frac 12(\omega^{+-}_{\bar\bullet} - \omega^{\bullet\bar\bullet}_{\bar\bullet})
+ \frac i2 A_{\bar\bullet} + \partial_{\bar\bullet}q^X{\omega_{X\,1}}^1
+ g A^I_{\bar\bullet}{P_{I\,1}}^1 - \sqrt 2 g S^{11} = 0\,,
\end{equation}
\begin{equation}
-\omega^{-\bullet}_{\bar\bullet} - \bar b\partial_{\bar\bullet}q^X{\omega_{X\,2}}^1
- \bar b g A^I_{\bar\bullet}{P_{I\,2}}^1 = 0\,, \label{gravbarbullet2}
\end{equation}
\begin{align}
-\bar b\,\omega^{+\bar\bullet}_{\bar\bullet} + &\partial_{\bar\bullet}q^X{\omega_{X\,1}}^2
+ g A^I_{\bar\bullet}{P_{I\,1}}^2 - \sqrt 2 g S^{21} \nonumber \\
&-\frac 1{\sqrt2}(F^{-I+-}-F^{-I\bullet\bar\bullet})
(\text{Im}\,{\cal N})_{IJ}Z^J e^{{\cal K}/2} = 0\,,
\end{align}
\vspace*{-0.8cm}
\begin{align}
-\partial_{\bar\bullet}\bar b - &\frac{\bar b}2(\omega^{\bullet\bar\bullet}_{\bar\bullet}
- \omega^{+-}_{\bar\bullet}) - \frac{i\bar b}2 A_{\bar\bullet} - \bar b
\partial_{\bar\bullet}q^X{\omega_{X\,2}}^2 \nonumber \\
&- \bar b g A^I_{\bar\bullet}{P_{I\,2}}^2
+ \sqrt2 F^{-I-\bullet}(\text{Im}\,{\cal N})_{IJ}Z^J e^{{\cal K}/2} = 0\,.
\label{gravbarbullet4}
\end{align}

\subsection{Geometry of spacetime}
\label{geom}

In order to obtain the spacetime geometry, we consider the spinor bilinears
\begin{equation}
V_{\mu\:\;j}^{\:i} = A(\epsilon^i, \Gamma_{\mu}\epsilon_j)\,,
\end{equation}
where the Majorana inner product $A$ is defined in \eqref{Majorana}.
The nonvanishing components are
\begin{equation}
V_{-\:\;1}^{\:1} = -\sqrt 2\,, \qquad V_{+\:\;2}^{\:2} = \sqrt 2 \bar b b\,, \qquad
V_{\bullet\:\;2}^{\:1} = \sqrt 2 b\,, \qquad V_{\bar\bullet\:\;1}^{\:2} =
\sqrt 2 \bar b\,.
\end{equation}
Note that $V_{\mu\:\;j}^{\:i} = V_{\mu\:\;i}^{\:j\;\ast}$, so that we can expand into
a basis of hermitian matrices,
\begin{equation}
V_{\mu\:\;j}^{\:i} = \frac 12 V_{\mu}\delta^i_{\;j} + \vec V_{\mu}\cdot
\vec{\sigma}_j^{\;\:i}\,.
\end{equation}
This yields for the trace part
\begin{equation}
V_{\mu}dx^{\mu} = \sqrt 2(|b|^2 E^+ - E^-)\,,
\end{equation}
while the nonzero components of $\vec V_{\mu}$ read
\begin{displaymath}
V^1_{\bullet} = \frac b{\sqrt 2}\,, \quad V^1_{\bar\bullet} = \frac{\bar b}{\sqrt 2}\,,
\quad V^2_{\bullet} = -\frac{ib}{\sqrt 2}\,, \quad V^2_{\bar\bullet} =
\frac{i\bar b}{\sqrt 2}\,, \quad V^3_+ = -\frac{\bar b b}{\sqrt 2}\,, \quad
V^3_- = -\frac 1{\sqrt 2}\,.
\end{displaymath}
Using the identities
\begin{displaymath}
{\omega_{X\,i}}^{j\,\ast} = -{\omega_{X\,j}}^i\,, \qquad
{P_{I\,i}}^{j\,\ast} = -{P_{I\,j}}^i\,, \qquad S^{ij} = S^{ji}\,,
\end{displaymath}
it is straightforward to shew that the linear system
\eqref{grav+1} - \eqref{gravbarbullet4}
implies the following constraints on the spin connection:
\begin{displaymath}
\omega^{+-}_+ = \partial_+\ln(\bar b b) = \partial_-(\bar b b)\,, \qquad
\omega^{+-}_- = 0\,, \qquad -\bar b b\,\omega^{+\bar\bullet}_- + \omega^{-\bar\bullet}_-
- \omega^{+-}_{\bullet} = 0\,,
\end{displaymath}
\begin{displaymath}
\bar b b\,\omega^{+-}_{\bullet} - \partial_{\bullet}(\bar b b) -
\omega^{-\bar\bullet}_+ + \bar b b\,\omega^{+\bar\bullet}_+ = 0\,, \qquad
-\bar b b\,\omega^{+\bar\bullet}_{\bullet} + \omega^{-\bar\bullet}_{\bullet} = 0\,,
\end{displaymath}
\begin{equation}
-\bar b b(\omega^{+\bar\bullet}_{\bar\bullet} + \omega^{+\bullet}_{\bullet}) +
\omega^{-\bullet}_{\bullet} + \omega^{-\bar\bullet}_{\bar\bullet} = 0\,.
\end{equation}
These are ten real equations, which are easily shown to be equivalent to
\begin{equation}
\partial_A V_B + \partial_B V_A - {\omega^C}_{B|A} V_C - {\omega^C}_{A|B} V_C = 0\,,
\end{equation}
which means that $V$ is Killing. Note that $V^2 = -4\bar b b$, so $V$ is timelike.
Moreover, one verifies that the system \eqref{grav+1} - \eqref{gravbarbullet4}
yields the relations
\begin{equation}
dV^x + \epsilon^{xyz}{\cal A}^y\wedge V^z = T^x\,, \label{dVx}
\end{equation}
with the gauged SU(2) connection
\begin{equation}
\vec{{\cal A}}_{\mu} = 2\partial_{\mu}q^X\vec{\omega}_X + 2g A^I_{\mu}\vec P_I\,,
\end{equation}
where we switched from SU(2) indices to vector quantities using the conventions
of appendix \ref{conv}. The torsion tensor\footnote{The reason for choosing this name
will be explained in appendix \ref{hol-base}.} $T^x$ can be written as
\begin{equation}
T^x = -\epsilon^{xyz}B^y\wedge V^z\,,
\end{equation}
with the one-form $B^y$ given by
\begin{displaymath}
B^1_+ = -2\sqrt 2 g\,\text{Im}(bS^{22})\,, \qquad B^1_- = -2\sqrt 2 g\,\text{Im}\left(
\frac{S^{11}}{\bar b}\right)\,, \qquad B^1_{\bullet} = -\frac{2\sqrt 2 g i}{\bar b}\,
\text{Re}(bS^{12})\,,
\end{displaymath}
\begin{displaymath}
B^2_+ = -2\sqrt 2 g\,\text{Re}(bS^{22})\,, \qquad B^2_- = 2\sqrt 2 g\,\text{Re}\left(
\frac{S^{11}}{\bar b}\right)\,, \qquad B^2_{\bullet} = -iB^1_{\bullet}\,,
\end{displaymath}
\begin{equation}
B^3_+ = -2\sqrt 2 g\,\text{Im}(bS^{12})\,, \qquad B^3_- = -\frac{B^3_+}{\bar b b}\,,
\qquad B^3_{\bullet} = \frac{\sqrt 2 g i}{\bar b}(bS^{22} - \bar b {\bar S}^{11})\,.
\label{B}
\end{equation}
Notice that we are free to include the torsion term in the SU(2) connection, by
rewriting \eqref{dVx} as
\begin{equation}
dV^x + \epsilon^{xyz}({\cal A}^y+B^y)\wedge V^z = 0\,, \label{dVxclosed}
\end{equation}
so that the forms $V^x$ are actually SU(2)-covariantly closed, similar to the
ungauged case \cite{Huebscher:2006mr}. If we define
\begin{displaymath}
{\cal A}_{\mu}^{\pm} \equiv {\cal A}^1_{\mu} \pm i {\cal A}^2_{\mu}\,,
\end{displaymath}
and similar for $B$, eqns.~\eqref{grav+3}, \eqref{grav-2}, \eqref{gravbullet3} and
\eqref{gravbarbullet2} can be cast into the form
\begin{displaymath}
b\,\omega^{+\bullet}_+ = -\frac i2({\cal A}^+_+ + B^+_+)\,, \qquad
\omega^{-\bar\bullet}_- = \frac{ib}2({\cal A}^-_- + B^-_-)\,,
\end{displaymath}
\begin{equation}
\omega^{-\bar\bullet}_{\bullet} = \frac{ib}2({\cal A}^-_{\bullet} + B^-_{\bullet})\,,
\qquad b\,\omega^{+\bullet}_{\bar\bullet} = -\frac i2({\cal A}^+_{\bar\bullet} +
B^+_{\bar\bullet})\,.
\end{equation}
These equations relate the SU(2) to the spin connection,
and tell us how the former is embedded into the latter. Such an embedding is
necessary for unbroken supersymmetry; it leads to a (partial) cancellation of
the SU(2) and spin connections in the gravitino supersymmetry transformation, and
generalizes the mechanism of \cite{Maldacena:2000mw,Klemm:2000nj}.

Let us now choose coordinates $(t,x^1,x^2,x^3)$ such that $V = \partial_t$.
The metric will then be independent of $t$. Note that
$\partial_t = \sqrt 2\,(|b|^2\partial_- - \partial_+)$. Making use of
\begin{displaymath}
{\omega_{X\,i}}^i = {P_{I\,i}}^i = 0\,,
\end{displaymath}
eqns.~\eqref{grav+1}, \eqref{grav+4}, \eqref{grav-1} and \eqref{grav-3} give
\begin{equation}
\partial_t\ln b = iA_t\,, \label{partial_tb}
\end{equation}
whose real part implies that $|b|$ is time-independent.
In terms of the vierbein $E^A_{\mu}$ the metric is given by
\begin{equation}
ds^2 = 2 E^+ E^- + 2 E^{\bullet} E^{\bar\bullet}\,, \label{metricE}
\end{equation}
where
\begin{displaymath}
E^+_{\mu} = \frac{V_{\mu} - 2V^3_{\mu}}{2\sqrt 2 |b|^2}\,, \qquad
E^-_{\mu} = -\frac{V_{\mu} + 2V^3_{\mu}}{2\sqrt 2}\,, \qquad
E^{\bullet}_{\mu} = \frac{V^1_{\mu}+iV^2_{\mu}}{\sqrt 2 b}\,, \qquad
E^{\bar\bullet}_{\mu} = \frac{V^1_{\mu}-iV^2_{\mu}}{\sqrt 2\bar b}\,.
\end{displaymath}
From $V^2 = -4|b|^2$ and $V = \partial_t$ as a vector we get
$V_t = -4|b|^2$, so that $V = - 4|b|^2 (dt + \sigma)$ as a one-form,
with $\sigma_t = 0$. Furthermore, $V^{\bullet} = 0$ yields
$E^{\bullet}_t = 0$ and thus $V^1_t = V^2_t = 0$. Since $V$ and $V^3$ are
orthogonal, $V^{\mu}V^3_{\mu} = 0$, also $V^3_t$ vanishes, and hence
$V^x_t=0$. The metric (\ref{metricE}) becomes thus
\begin{equation}
ds^2 = - 4|b|^2 (dt + \sigma)^2 + |b|^{-2}\delta_{xy}V^x V^y\,. \label{metric}
\end{equation}
In order to proceed one would like to choose the gauge
${\cal A}^x_t + B^x_t = 0$, which reduces to the choice made in \cite{Huebscher:2006mr}
for $g\to 0$. Then the SU(2)-covariant closure of the $V^x$
(eq.~\eqref{dVxclosed}) states that the SU(2) connection ${\cal A}+B$ and
the $V^x$ are time-independent. Eq.~\eqref{dVxclosed} can then be interpreted
as Cartan's first structure equation on the three-dimensional base space.
One therefore has to show that the above gauge is always possible. Let us
at this point restrict to the case without hypers and no gauging
of special K\"ahler isometries ($k^{\alpha}_I=0$). The inclusion of hypermultiplets will be
studied in a forthcoming publication \cite{chkmmovz}. This leaves two possible
solutions for the moment maps \cite{Vambroes}, namely SU(2) or U(1)
Fayet-Iliopoulos (FI) terms.
We shall consider here the latter, which satisfy \eqref{FI}, where
$e^x = \delta^x_3$ without loss of generality\footnote{$e^x = \delta^x_3$
can always be achieved by a global SU(2)
rotation (which is a symmetry of the theory).}. One has then
\begin{displaymath}
{P_{I\,1}}^1 = -{P_{I\,2}}^2 = i\,\xi_I\,, \qquad {P_{I\,1}}^2 = {P_{I\,2}}^1 = 0\,,
\end{displaymath}
\begin{equation}
S^{12} = S^{21} = i\,\xi_I Z^I e^{{\cal K}/2}\,, \qquad S^{11} = S^{22} = 0\,,
\label{S-FI}
\end{equation}
as well as
\begin{equation}
{\cal A}^1_{\mu} = {\cal A}^2_{\mu} = 0\,, \qquad
{\cal A}^3_{\mu} = 2g A^I_{\mu}\xi_I\,. \label{red-hol}
\end{equation}
From \eqref{dVxclosed} one obtains $dV^3=0$, like in minimal gauged
supergravity \cite{Caldarelli:2003pb,Cacciatori:2007vn}. If we choose the
gauge ${\cal A}^3_t + B^3_t = 0$, the one-forms $V^x$ will be time-independent.
Note that the U(1) gauge transformation \eqref{U(1)transf} necessary to achieve
this does not spoil our choice of representatives: As discussed in section \ref{orbits},
the phase factors acquired by the Killing spinors $\epsilon^i$ (eq.~\eqref{transf_eps})
can be eliminated by a subsequent Spin(3,1) transformation.
The above gauge condition implies
\begin{equation}
 A^I_t\xi_I = -4\,\text{Im}(bS^{12})\,, \label{gauge-cond}
\end{equation}
and is left invariant by transformations \eqref{U(1)transf} with time-independent
$\xi_I\alpha^I$. As the SU(2) connection ${\cal A}+B$ and the $V^x$ do not
depend on $t$, one can regard \eqref{dVxclosed} as Cartan's first structure
equation on the three-dimensional base manifold with metric $\delta_{xy}V^xV^y$.

Next we consider the equations coming from the gaugino variation. Using
\begin{displaymath}
N^{\alpha}_{11} = N^{\alpha}_{22} = 0\,, \qquad N^{\alpha}_{12} =
N^{\alpha}_{21} = -2i\,\xi_I e^{{\cal K}/2}{\cal D}_{\bar\beta}{\bar Z}^I
g^{\alpha\bar\beta}\,,
\end{displaymath}
and ${\cal D}_{\mu}z^{\alpha} = \partial_{\mu}z^{\alpha}$, eqns.~\eqref{G1} and
\eqref{G3} yield
\begin{displaymath}
\partial_t z^{\alpha} = 0\,,
\end{displaymath}
i.~e.~, the scalar fields are time-independent. Choosing the constants $C_{I,JK}=0$
and taking into account that the structure constants ${f_{IJ}}^K$
vanish also, eqns.~\eqref{P0I} and
\eqref{TIV} imply for the moment map function $P^0_I=0$.
But then from \eqref{KaehlerU1} one has for the K\"ahler
U(1)
\begin{displaymath}
A_t = -\frac i2(\partial_{\alpha}{\cal K}\partial_t z^{\alpha} - \partial_{\bar\alpha}
{\cal K}\partial_t{\bar z}^{\bar\alpha}) = 0\,.
\end{displaymath}
Plugging this into \eqref{partial_tb} gives $\partial_t b = 0$, hence $b$ is
time-independent as well.

Notice that the system \eqref{grav+1} - \eqref{gravbarbullet4} allows to
express the linear combinations $A^I\xi_I$ and
$F^{-I}(\text{Im}\,{\cal N})_{IJ}Z^J$ of the gauge potentials and fluxes
in terms of the spin connection, the K\"ahler
U(1), the linear combination of scalars $Z^I\xi_I$ and the function $b$,
\begin{align}
ig A^I_+\xi_I = &\frac 12\omega^{\bullet\bar\bullet}_+ + \frac i2 A_+
-\frac 12\partial_+\ln\frac b{\bar b}\,, \qquad
ig A^I_-\xi_I = \frac 12\omega^{\bullet\bar\bullet}_- - \frac i2 A_-\,,
\nonumber \\
&ig A^I_{\bullet}\xi_I = \frac 12(\omega^{+-}_{\bullet} + \omega^{\bullet
\bar\bullet}_{\bullet}) - \frac i2 A_{\bullet}\,, \label{xiA}
\end{align}
\begin{align}
F^{-I+\bar\bullet}&(\text{Im}\,{\cal N})_{IJ}Z^J e^{{\cal K}/2} = \frac{\bar b}{\sqrt 2}
\omega^{+\bar\bullet}_-\,, \qquad
F^{-I-\bullet}(\text{Im}\,{\cal N})_{IJ}Z^J e^{{\cal K}/2} = -\frac 1{\sqrt 2 b}
\omega^{-\bullet}_+\,, \nonumber \\
&(F^{-I+-} - F^{-I\bullet\bar\bullet})(\text{Im}\,{\cal N})_{IJ}Z^J e^{{\cal K}/2}
= -\sqrt 2\bar b\,\omega^{+\bar\bullet}_{\bar\bullet} - 2ig\,\xi_I e^{{\cal K}/2}Z^I\,.
\label{FNZ}
\end{align}
As the $(n_V+1)\times(n_V+1)$ matrix $(X^I, {\cal D}_{\bar\alpha}{\bar X}^I)$
is invertible \cite{Vambroes}, \eqref{FNZ} together with the gaugino
equations \eqref{G1}-\eqref{G4} determine uniquely the fluxes $F^{-I}$,
with the result\footnote{To get this, one has to use \eqref{inverse}.}
\begin{align}
F^{-I+\bar\bullet}&= \frac{\sqrt 2}b{\bar X}^I(\partial_{\bullet}\ln\bar b+i
A_{\bullet})+\frac{\sqrt 2}{\bar b}{\cal D}_{\alpha}X^I\partial_{\bullet}
z^{\alpha}\,, \nonumber \\
F^{-I-\bullet}&= -\sqrt 2\,\bar b{\bar X}^I(\partial_{\bar\bullet}\ln\bar b+i
A_{\bar\bullet})-\sqrt 2\,b{\cal D}_{\alpha}X^I\partial_{\bar\bullet}
z^{\alpha}\,, \label{fluxes} \\
F^{-I+-}-F^{-I\bullet\bar\bullet}&=\frac{2\sqrt2}b \bar X^I(\partial_+\ln\bar b
+iA_+) + \frac{2\sqrt 2}{\bar b}{\cal D}_{\alpha}X^I\partial_+z^{\alpha}
+2ig\,\xi_J(\text{Im}\,{\cal N})^{-1|IJ}\,. \nonumber
\end{align}
Moreover, antiselfduality implies that
\begin{displaymath}
F^{-I+\bullet} = F^{-I-\bar\bullet} = F^{-I+-}+F^{-I\bullet\bar\bullet} = 0\,.
\end{displaymath}
With \eqref{fluxes}, all gaugino equations are satisfied.

Furthermore, the system \eqref{grav+1} - \eqref{gravbarbullet4} determines
almost all components of the spin connection (with the exception of
$\omega^{\bullet\bar\bullet}$) in terms of $A_{\mu}$, $Z^I\xi_I$, the
function $b$ and its spacetime derivatives,
\begin{displaymath}
\omega^{+-}_+ = \partial_+\ln(\bar b b)\,, \qquad
\omega^{+-}_- = 0\,, \qquad \omega^{+-}_{\bullet} =
\partial_{\bullet}\ln\bar b + i A_{\bullet}\,,
\end{displaymath}
\begin{displaymath}
\omega^{+\bullet}_+ = \omega^{+\bullet}_{\bar\bullet} = 0\,, \qquad
\omega^{+\bullet}_- = -\frac 1{\bar b b}(\partial_{\bar\bullet}\ln b -
i A_{\bar\bullet})\,,
\end{displaymath}
\begin{displaymath}
\omega^{+\bullet}_{\bullet} = \partial_-\ln b - i A_- +
\frac{2\sqrt 2 i g}b\,\xi_I e^{{\cal K}/2}{\bar Z}^I\,,
\end{displaymath}
\begin{displaymath}
\omega^{-\bullet}_+ = - b\,\partial_{\bar\bullet}\bar b - i\bar b b A_{\bar\bullet}\,,
\qquad \omega^{-\bullet}_- = \omega^{-\bullet}_{\bar\bullet} = 0\,,
\end{displaymath}
\begin{equation}
\omega^{-\bullet}_{\bullet} = \partial_+\ln\bar b + i A_+ -
2\sqrt 2 gbi\,\xi_I e^{{\cal K}/2}Z^I\,. \label{spinconn}
\end{equation}
From the gauge condition \eqref{gauge-cond} we obtain one more component,
namely
\begin{equation}
\omega^{\bullet\bar\bullet}_t = \sqrt 2(|b|^2\omega^{\bullet\bar\bullet}_-
-\omega^{\bullet\bar\bullet}_+) = -\sqrt 2\,\partial_+\ln\frac b{\bar b}
+2\sqrt 2 iA_+ - 4ig\xi_I(bX^I+\bar b\bar X^I)\,.
\end{equation}
The next step is to impose vanishing spacetime torsion,
\begin{displaymath}
\partial_{\mu} E^A_{\nu} - \partial_{\nu} E^A_{\mu} + \omega^A_{\mu B}
E^B_{\nu} - \omega^A_{\nu B} E^B_{\mu} = 0\,.
\end{displaymath}
One finds that most of these equations are already identically
satisfied, while the remaining ones yield (using the expressions
(\ref{spinconn}) for the spin connection)
\begin{equation}
d\sigma + \zeta^x\epsilon^{xyz} V^y\wedge V^z = 0\,, \label{dsigma}
\end{equation}
where the (real) SU(2) vector $\zeta^x$ is defined as
\begin{align}
&\zeta^1 + i\zeta^2 = -\frac 1{\sqrt 2\,{\bar b}^2 b}\left(\frac i2\partial_{\bar\bullet}
\ln\frac b{\bar b} + A_{\bar\bullet}\right)\,, \nonumber \\
\zeta^3 = \frac 1{\sqrt 2 |b|^2}&\left[\frac i2\partial_-\ln\frac b{\bar b} + A_-
- \sqrt 2 g\,\xi_I e^{{\cal K}/2}\left(\frac{{\bar Z}^I}b + \frac{Z^I}{\bar b}
\right)\right]\,.
\end{align}
We already noted that $dV^3=0$, hence there exists a function $z$ such that
$V^3=dz$ locally. Since $V^3_t=0$, $z$ must be time-independent. Let us use
$z$ as one of the coordinates $x^1, x^2, x^3$, say $z=x^3$. The remaining
spatial coordinates will be denoted by late small latin indices $m,n,\ldots$,
i.~e.~, $x^m=x^1,x^2$, while capital late latin indices $M,N,\ldots=1,2$ refer
to the corresponding tangent space. One can eliminate the components $V^M_z$
by a diffeomorphism
\begin{displaymath}
x^m = x^m({x'}^n, z)\,,
\end{displaymath}
with
\begin{displaymath}
V^M_m\frac{\partial x^m}{\partial z} = -V^M_z\,.
\end{displaymath}
As the matrix $V^M_m$ is invertible\footnote{This follows from
$\sqrt{-g}=2\det(V^M_m)/|b|^2$.}, one can always solve for
$\partial x^m/\partial z$. Notice that the metric \eqref{metric} is
invariant under
\begin{displaymath}
t \to t + \chi(x^m, z)\,, \qquad \sigma \to \sigma - d\chi\,,
\end{displaymath}
for an arbitrary function $\chi(x^m,z)$. This second gauge freedom can be
used to set $\sigma_z=0$. Thus, without loss of generality, we can take
$\sigma=\sigma_m dx^m$, and the metric \eqref{metric} becomes
\begin{equation}
ds^2 = -4|b|^2(dt+\sigma_m dx^m)^2 + |b|^{-2}\left(dz^2 + \delta_{MN}V^MV^N\right)\,.
\label{metric1}
\end{equation}
The solution of the Cartan structure equation \eqref{dVx} is then given by
\begin{align}
&V^1_m+iV^2_m = (\hat V^1_m+i\hat V^2_m)\left(\frac b{\bar b}\right)^{\frac 12}
\exp\Phi\,, \\
&\omega^{\bullet\bar\bullet}_{\bullet} = \frac{|b|}{\sqrt 2}e^{-\Phi}(\hat V^m_1
-i\hat V^m_2)\left[-i\hat\omega_m + \partial_m(\bar\Phi-\ln |b|)\right]\,,
\end{align}
where $\hat V^M_m$ denote integration "constants" depending only on $x^n$
but not on $z$, $\hat V^m_M$ is the corresponding inverse zweibein,
$\Phi$ is a complex function defined by
\begin{equation}
\partial_z\Phi = 2ig\xi_I\left(\frac{\bar X^I}b - \frac{X^I}{\bar b}\right)
-\omega^{\bullet\bar\bullet}_z\,, \label{Phi}
\end{equation}
and $\hat\omega\equiv\hat\omega^{12}$ is the spin connection following from
the zweibein $\hat V^M$. At this point it is convenient to use the residual
U(1) gauge freedom of a combined local Lorentz and electromagnetic gauge
transformation to eliminate $\omega^{\bullet\bar\bullet}_z$. This is accomplished
by the transformation \eqref{residualU1}, with
\begin{displaymath}
\psi = \frac i2\int dz\omega^{\bullet\bar\bullet}_z\,.
\end{displaymath}
Note that $\psi$ is real, as it must be. Moreover, as $\psi$ is time-independent,
this does not spoil the gauge choice \eqref{gauge-cond}. With
$\omega^{\bullet\bar\bullet}_z=0$, $\Phi$ is real.
In what follows, we shall introduce complex coordinates $w=x^1+ix^2$,
$\bar w = x^1-ix^2$, and choose the conformal gauge for the two-metric
$\delta_{MN}\hat V^M_m\hat V^N_n$, i.~e.~,
\begin{equation}
\delta_{MN}\hat V^M_m\hat V^N_n = e^{2\gamma}dw d\bar w\,,
\end{equation}
where $\gamma = \gamma(w,\bar w)$. From \eqref{Phi} it is clear that
$\Phi$ is defined only up to an arbitrary function of
$w,\bar w$. This allows to absorb $\gamma$ into $\Phi$, so one can take
$\gamma=0$ without loss of generality. Then the metric \eqref{metric1}
simplifies to
\begin{equation}
ds^2 = -4|b|^2(dt+\sigma)^2 + |b|^{-2}\left(dz^2 + e^{2\Phi}dw d\bar w\right)\,,
\label{final-metric}
\end{equation}
with $\sigma=\sigma_w dw+\sigma_{\bar w}d\bar w$.

Defining the symplectic vector
\begin{equation}
{\cal I} = \text{Im}\left({\cal V}/\bar b\right)\,,
\end{equation}
where $\cal V$ is given in \eqref{sympl-vec}, eq.~\eqref{dsigma} can be cast into
the form
\begin{equation}
d\sigma + 2\,\star^{(3)}\!\langle{\cal I}\,,d{\cal I}\rangle - \frac{ig}{|b|^2}\xi_I\left(\frac{\bar X^I}b
+\frac{X^I}{\bar b}\right)e^{2\Phi}dw\wedge d\bar w=0\,. \label{dsigma-sympl}
\end{equation}
Here $\star^{(3)}$ is the Hodge star on the three-dimensional base with dreibein
$V^x$. In the ungauged case $g=0$, \eqref{dsigma-sympl} reduces correctly to the expression
given in \cite{Huebscher:2006mr}.

All that remains to be done at this point is to impose the Bianchi identities and the
Maxwell equations, which read respectively
\begin{equation}
dF^I=0\,, \qquad d\,\text{Re}\,G^+_I=0\,,
\end{equation}
where $G^{\pm}_I={\cal N}_{IJ}F^{\pm J}$. One finds that the Bianchi identities are
equivalent to
\begin{eqnarray}
&&\qquad 4\partial\bar\partial\left(\frac{X^I}{\bar b}-\frac{\bar X^I}b\right) + \partial_z\left[e^{2\Phi}\partial_z
\left(\frac{X^I}{\bar b}-\frac{\bar X_I}b\right)\right]  \label{bianchi} \\
&&-2ig\xi_J\partial_z\left\{e^{2\Phi}\left[|b|^{-2}(\text{Im}\,{\cal N})^{-1|IJ}
+ 2\left(\frac{X^I}{\bar b}+\frac{\bar X^I}b\right)\left(\frac{X^J}{\bar b}+\frac{\bar X^J}b\right)\right]\right\}= 0\,,
\nonumber
\end{eqnarray}
while the Maxwell equations yield
\begin{eqnarray}
&&\qquad 4\partial\bar\partial\left(\frac{F_I}{\bar b}-\frac{\bar F_I}b\right) + \partial_z\left[e^{2\Phi}\partial_z
\left(\frac{F_I}{\bar b}-\frac{\bar F_I}b\right)\right] \nonumber \\
&&-2ig\xi_J\partial_z\left\{e^{2\Phi}\left[|b|^{-2}\text{Re}\,{\cal N}_{IL}(\text{Im}\,{\cal N})^{-1|JL}
+ 2\left(\frac{F_I}{\bar b}+\frac{\bar F_I}b\right)\left(\frac{X^J}{\bar b}+\frac{\bar X^J}b\right)\right]\right\}
\nonumber \\
&&-8ig\xi_I e^{2\Phi}\left[\langle {\cal I}\,,\partial_z {\cal I}\rangle-\frac g{|b|^2}\xi_J\left(\frac{X^J}{\bar b}
+\frac{\bar X^J}b\right)\right] = 0\,. \label{maxwell}
\end{eqnarray}
Here we defined $\partial=\partial_w$, $\bar\partial=\partial_{\bar w}$. Note that imposing $dF^I=0$
is actually not sufficient; we must also ensure that $\xi_I F^I=\xi_I dA^I$, because the linear
combination $\xi_I A^I$ is determined by the Killing spinor equations (cf.~eq.~\eqref{xiA}).
This gives the additional condition
\begin{equation}
2\partial\bar\partial\Phi=g e^{2\Phi}\left[i\xi_I\partial_z\left(\frac{X^I}{\bar b}-\frac{\bar X^I}b\right)
+2g|b|^{-2}\xi_I\xi_J(\text{Im}\,{\cal N})^{-1|IJ}+4g\left(\frac{\xi_I X^I}{\bar b}+\frac{\xi_I \bar X^I}b
\right)^2\right]\,, \label{Delta-Phi}
\end{equation}
which is slightly stronger than the contraction of \eqref{bianchi} with $\xi_I$\footnote{Contracting
\eqref{bianchi} with $\xi_I$ and using \eqref{Phi}, one gets the derivative of \eqref{Delta-Phi}
with respect to $z$.}.

Finally, note that the integrability condition for \eqref{dsigma-sympl}, namely
\begin{equation}
2\langle{\cal I}\,, \Delta^{(3)}{\cal I}\rangle = \star^{(3)} d\left[ig|b|^{-2}\xi_I\left(\frac{X^I}{\bar b}
+\frac{\bar X^I}b\right)e^{2\Phi}dw\wedge d\bar w\right]\,,
\end{equation}
where $\Delta^{(3)}$ denotes the Laplacian on the three-dimensional base manifold,
follow from the Bianchi identities and the Maxwell equations. One can show this by using
some relations of special K\"ahler geometry.

In conclusion, the functions $b$ and $\Phi$ together with the scalar fields are determined
by the equations \eqref{Phi}, \eqref{bianchi}, \eqref{maxwell} and \eqref{Delta-Phi}.
Then, the shift vector $\sigma$ is obtained from \eqref{dsigma-sympl} and the metric
is given by \eqref{final-metric}. The gauge fields can be read off from \eqref{fluxes}, which
can be rewritten as
\begin{eqnarray}
F^I&=&2(dt+\sigma)\wedge d\left[bX^I+\bar b\bar X^I\right]+|b|^{-2}dz\wedge d\bar w
\left[\bar X^I(\bar\partial\bar b+iA_{\bar w}\bar b)+({\cal D}_{\alpha}X^I)b\bar\partial z^{\alpha}-
\right. \nonumber \\
&&\left. X^I(\bar\partial b-iA_{\bar w}b)-({\cal D}_{\bar\alpha}\bar X^I)\bar b\bar\partial\bar z^{\bar\alpha}
\right]-|b|^{-2}dz\wedge dw\left[\bar X^I(\partial\bar b+iA_w\bar b)+\right. \nonumber \\
&&\left.({\cal D}_{\alpha}X^I)b\partial z^{\alpha}-X^I(\partial b-iA_w b)-({\cal D}_{\bar\alpha}\bar X^I)
\bar b\partial\bar z^{\bar\alpha}\right]- \nonumber \\
&&\frac 12|b|^{-2}e^{2\Phi}dw\wedge d\bar w\left[\bar X^I(\partial_z\bar b+iA_z\bar b)+({\cal D}_{\alpha}
X^I)b\partial_z z^{\alpha}-X^I(\partial_z b-iA_z b)- \right.\nonumber \\
&&\left.({\cal D}_{\bar\alpha}\bar X^I)\bar b\partial_z\bar z^{\bar\alpha}-2ig\xi_J
(\text{Im}\,{\cal N})^{-1|IJ}\right]\,. \label{final-fluxes}
\end{eqnarray}
Notice that, in the timelike case, the vanishing of the supersymmetry
variations, together with the Bianchi identities and the Maxwell
equations, imply all the equations of
motion. This is shown in appendix \ref{BPS-eom}.

In a forthcoming paper \cite{attractors} we shall consider various models (specified by a certain
prepotential), and give explicit solutions of the above equations that represent supersymmetric
AdS black holes with nontrivial scalar fields turned on.

\section{Final remarks}
\label{finalrem}

In this paper, we applied spinorial geometry techniques to classify all
supersymmetric solutions of ${\cal N}=2$ gauged supergravity in four dimensions
coupled to abelian vector multiplets. Our results can be used to construct new BPS black holes
in AdS$_4$ with nonconstant scalars. Such solutions are, to the best of our knowledge, unknown
up to now, and would be important to study the attractor mechanism in
AdS \cite{Bellucci:2008cb}. This will be the subject of a future publication \cite{attractors}.

Possible extensions of our work could be to impose the existence of more than one Killing
spinor and to determine how this constrains further the geometry of supersymmetric backgrounds,
as was done in the minimal case in \cite{Cacciatori:2007vn}. It would also be interesting to see
if nontrivial preons (i.e., solutions with nearly maximal supersymmetry that are not simply quotients
of AdS) exist in matter-coupled gauged supergravity.

In refs.~\cite{Huebscher:2007hj, Meessen:2008kb}, the ${\cal N}=2$, $D=4$ theory coupled to
non-abelian vector multiplets with a gauge group that includes an SU(2) factor was considered,
and various supersymmetric solutions, such as embeddings of the 't Hooft-Polyakov monopole
and extremal black holes were obtained. These geometries are asymptotically flat,
and it would be very interesting to find similar solutions in the asymptotically AdS case,
for instance in ${\cal N}=2$ supergravity where the full SU(2) R-symmetry is gauged,
which can induce a negative cosmological constant. There are only very few analytically known
Einstein-Yang-Mills black holes, and to dispose of more solutions would of course be helpful
in probing the validity of the no-hair conjecture. Of particular relevance in this context are black holes
with AdS asymptotics, which were recently argued to require an infinite number of parameters for
their description \cite{Baxter:2007at}. This is one of the reasons that make it desirable to systematically
classify all supersymmetric backgrounds of ${\cal N}=2$, $D=4$ supergravity with general gauging.
Work in this direction is in progress \cite{chkmmovz}.

\acknowledgments

This work was partially supported by INFN, PRIN prot.~2005024045-002 and
by the European Commission program MRTN-CT-2004-005104. We would like to thank
M.~M.~Caldarelli, D.~Roest, A.~Van Proeyen and especially M.~H\"ubscher,
P.~Meessen, T.~Ort\'{\i}n and S.~Vaul\`a for useful discussions. D.~K.~,
D.~S.~M.~and E.~Z.~wish to thank the Instituto de F\'{\i}sica Te\'orica
UAM/CSIC for hospitality.

\normalsize

\appendix

\section{Conventions}
\label{conv}

We use the notations and conventions of \cite{Vambroes}, which are briefly summarized here. More
details can be found in appendix A of \cite{Vambroes}.

The signature is mostly plus. Late greek letters $\mu,\nu,\ldots$ are curved spacetime indices,
while early latin letters $a,b,\ldots=0,\ldots,3$ and $A,B,\ldots=+,-,\bullet,\bar\bullet$ refer to the
corresponding tangent space, cf.~also appendix \ref{spin-forms}.

Self-dual and anti-self-dual field strengths are defined by
\begin{equation}
F^{\pm I}_{ab} = \frac 12(F^I_{ab}\pm \tilde F^I_{ab})\,,
\qquad \tilde F^I_{ab} \equiv -\frac i2\epsilon_{abcd}F^{I cd}\,,
\end{equation}
where $\epsilon_{0123}=1$, $\epsilon^{0123}=-1$. We also introduce
\begin{equation}
\epsilon^{\mu\nu\rho\sigma} = e\,e^{\mu}_a e^{\nu}_b e^{\rho}_c e^{\sigma}_d\epsilon^{abcd}\,.
\end{equation}
The $p$-form associated to an antisymmetric tensor $T_{\mu_1\ldots\mu_p}$ is
\begin{equation}
T = \frac 1{p!}T_{\mu_1\ldots\mu_p}dx^{\mu_1}\wedge\ldots\wedge dx^{\mu_p}\,, \label{T}
\end{equation}
and the exterior derivative acts as\footnote{Our definitions for $p$-forms, eq.~\eqref{T}, and for
exterior derivatives, eq.~\eqref{dT}, are the only points where our conventions differ from those
of \cite{Vambroes}.}
\begin{equation}
dT = \frac 1{p!}T_{\mu_1\ldots\mu_p,\nu}dx^{\nu}\wedge dx^{\mu_1}\wedge\ldots\wedge dx^{\mu_p}\,.
\label{dT}
\end{equation}
Antisymmetric tensors are often contracted with $\Gamma$-matrices as in
$\Gamma\cdot F\equiv \Gamma^{ab}F_{ab}$.

$i,j,\ldots=1,2$ are SU(2) indices, whose raising and lowering is done by complex conjugation.
The Levi-Civita $\epsilon^{ij}$ has the property
\begin{equation}
\epsilon_{ij}\epsilon^{jk} = -{\delta_i}^k\,,
\end{equation}
where in principle $\epsilon^{ij}$ is the complex conjugate of $\epsilon_{ij}$, but we can choose
$\epsilon=i\sigma_2$, such that
\begin{equation}
\epsilon_{12} = \epsilon^{12} = 1\,.
\end{equation}
The Pauli matrices ${\sigma_{xi}}^j$ ($x=1,2,3$) are given by
\begin{equation}
\sigma_1 = \left(\begin{array}{cc} 0 & 1 \\ 1 & 0\end{array}\right)\,, \qquad
\sigma_2 = \left(\begin{array}{cc} 0 & -i \\ i & 0\end{array}\right)\,, \qquad
\sigma_3 = \left(\begin{array}{cc} 1 & 0 \\ 0 & -1\end{array}\right)\,.
\end{equation}
They allow to switch from SU(2) indices to vector quantities using the convention
\begin{equation}
{A_i}^j \equiv i\vec A\cdot\vec\sigma_i^{\;\;j}\,.
\end{equation}
At various places in the main text we use $\sigma$-matrices with only lower or upper indices, defined by
\begin{equation}
\vec\sigma_{ij}\equiv \vec\sigma_i^{\;\;k}\epsilon_{kj}\,, \qquad i\vec\sigma^{ij} = (i\vec\sigma_{ij})^{\ast}\,.
\end{equation}
Notice that both $\vec\sigma_{ij}$ and $\vec\sigma^{ij}$ are symmetric.

Spinors carrying an index $i$ are chiral, e.g. for the supersymmetry parameter one has
\begin{equation}
\Gamma_5\epsilon^i = \epsilon^i\,, \qquad \Gamma_5\epsilon_i = -\epsilon_i\,,
\label{Gamma5-eps}
\end{equation}
and the same holds for the gravitino $\psi^i_{\mu}$. Note however that for some spinors, the upper index
denotes negative chirality rather than positive chirality, for instance the gauginos obey
\begin{equation}
\Gamma_5\lambda^{\alpha i} = -\lambda^{\alpha i}\,, \qquad \Gamma_5\lambda^{\alpha}_i =
\lambda^{\alpha}_i\,,
\end{equation}
as is also evident from the supersymmetry transformations. The charge conjugate of a spinor $\chi$
is
\begin{equation}
\chi^C = \Gamma_0 C^{-1}\chi^{\ast}\,,
\end{equation}
with the charge conjugation matrix $C$. Majorana spinors are defined by $\chi=\chi^C$,
and chiral spinors obey $\chi^C_i=\chi^i$.

\section{Spinors and forms}
\label{spin-forms}

In this appendix, we summarize the essential information needed to realize
the spinors of Spin(3,1) in terms of forms. For more details, we refer
to \cite{Lawson:1998yr}.
Let $V = \bR^{3,1}$ be a real vector space equipped with the Lorentzian inner
product $\langle\cdot,\cdot\rangle$. Introduce an orthonormal basis $e_1, e_2, e_3, e_0$,
where $e_0$ is along the time direction, and consider the subspace $U$
spanned by the first two basis vectors $e_1, e_2$. The space of Dirac spinors
is $\Delta_c = \Lambda^{\ast}(U\otimes \bC)$, with basis
$1, e_1, e_2, e_{12} = e_1 \wedge e_2$.
The gamma matrices are represented on $\Delta_c$ as
\eqn
\Gamma_{0}\eta&=&-e_2\wedge\eta+e_2\rfloor\eta\,, \qquad
\Gamma_{1}\eta=e_1\wedge\eta+e_1\rfloor\eta\,, \nonumber \\
\Gamma_{2}\eta&=&e_2\wedge\eta+e_2\rfloor\eta\,, \qquad
\Gamma_{3}\eta=ie_1\wedge\eta-ie_1\rfloor\eta\,,
\feqn
where
\begin{displaymath}
\eta = \frac 1{k!}\eta_{j_1\ldots j_k} e_{j_1}\wedge\ldots\wedge e_{j_k}
\end{displaymath}
is a $k$-form and
\begin{displaymath}
e_i \rfloor \eta = \frac 1{(k-1)!}\eta_{ij_1\ldots j_{k-1}} e_{j_1}\wedge\ldots
                  \wedge e_{j_{k-1}}\,.
\end{displaymath}
One easily checks that this representation of the gamma matrices satisfies
the Clifford algebra relations $\{\Gamma_a, \Gamma_b\} = 2\eta_{ab}$.
The parity matrix is defined by $\Gamma_5 = i\Gamma_0\Gamma_1\Gamma_2\Gamma_3$,
and one finds that the even forms $1, e_{12}$ have positive chirality,
$\Gamma_5\eta = \eta$, while the odd forms $e_1, e_2$ have negative chirality,
$\Gamma_5\eta = -\eta$, so that $\Delta_c$ decomposes into two complex chiral
Weyl representations $\Delta_c^+ = \Lambda^{\mathrm{even}}(U\otimes \bC)$ and
$\Delta_c^- = \Lambda^{\mathrm{odd}}(U\otimes \bC)$. Note that Spin(3,1) is
isomorphic to SL$(2,\bC)$, which acts with the fundamental representation on the
positive chirality Weyl spinors.\\
Let us define the auxiliary inner product
\begin{equation}
\langle\sum_{i=1}^2 \alpha_i e_i, \sum_{j=1}^2 \beta_j e_j\rangle = \sum_{i=1}^2
\alpha_i^{\ast}\beta_i
\end{equation}
on $U\otimes \bC$, and then extend it to $\Delta_c$. The Spin(3,1) invariant
Dirac inner product is then given by
\begin{equation}
D(\eta, \theta) = \langle\Gamma_0\eta, \theta\rangle\,.
\end{equation}
The Majorana inner product that we use is\footnote{It is known that on
even-dimensional manifolds there are two Spin invariant Majorana inner
products. The other possibility, based on $C=i\Gamma_{03}$, was used
in \cite{Grover:2006wy}.}
\begin{equation}
A(\eta, \theta) = \langle C\eta^{\ast}, \theta\rangle\,, \label{Majorana}
\end{equation}
with the charge conjugation matrix $C=\Gamma_{12}$. Using the identities
\begin{equation}
\Gamma_a^{\ast} = -C\Gamma_0\Gamma_a\Gamma_0 C^{-1}\,, \qquad
\Gamma_a^T = -C\Gamma_a C^{-1}\,,
\end{equation}
it is easy to show that \eqref{Majorana} is Spin(3,1) invariant as well.

The charge conjugation matrix $C$ acts on the basis elements as
\begin{equation}
C 1 = e_{12}\,, \quad C e_{12} = -1\,, \quad C e_1 = -e_2\,, \quad C e_2 = e_1\,.
\end{equation}

In many applications it is convenient to use a basis in which the
gamma matrices act like creation and annihilation operators, given
by \eqn
\Gamma_{+}\eta\equiv\frac1{\sqrt2}\left(\Gamma_{2}+\Gamma_{0}\right)\eta
&=&\sqrt2\,e_{2}\rfloor\eta\,, \qquad
\Gamma_{-}\eta\equiv\frac1{\sqrt2}\left(\Gamma_{2}-\Gamma_{0}\right)\eta
=\sqrt2\,e_{2}\wedge\eta\,, \nonumber \\
\Gamma_{\bullet}\eta\equiv\frac1{\sqrt2}\left(\Gamma_{1}-i\Gamma_{3}\right)\eta
&=&\sqrt2\,e_{1}\wedge\eta\,, \qquad
\Gamma_{\bar\bullet}\eta\equiv\frac1{\sqrt2}\left(\Gamma_{1}+i\Gamma_{3}\right)\eta
=\sqrt2\,e_{1}\rfloor\eta\,.
\feqn
The Clifford algebra relations in this  basis are $\{\Gamma_A,\Gamma_B\} = 2\eta_{AB}$,
where $A,B,\ldots = +,-,\bullet,\bar\bullet$ and the nonvanishing components of
the tangent space metric read
$\eta_{+-} = \eta_{-+} = \eta_{\bullet\bar\bullet} = \eta_{\bar\bullet\bullet} = 1$.
The spinor 1 is a Clifford vacuum, $\Gamma_{+}1 = \Gamma_{\bar\bullet}1 = 0$,
and the representation $\Delta_c$ can be constructed by acting on 1 with the
creation operators $\Gamma^+ = \Gamma_-, \Gamma^{\bar\bullet} = \Gamma_{\bullet}$,
so that any spinor can be written as
\begin{displaymath}
\eta = \sum_{k=0}^2 \frac 1{k!}\phi_{{\bar a}_1\ldots {\bar a}_k}\Gamma^{{\bar a}_1
\ldots {\bar a}_k}1\,, \qquad \bar a = +,\bar\bullet\,.
\end{displaymath}
The action of the Gamma matrices and the Lorentz generators $\Gamma_{AB}$ is
summarized in table \ref{tab:gamma}.

\begin{table}[ht]
\begin{center}
\begin{tabular}{|c||c|c|c|c|}
\hline
& 1 & $e_{1}$ & $e_{2}$ & $e_{1}\wedge e_{2}$\\
\hline\hline
$\Gamma_{+}$ & 0 & 0 & $\sqrt2$ & $-\sqrt2e_{1}$\\
\hline
$\Gamma_{-}$ & $\sqrt2e_{2}$ & $-\sqrt2e_{1}\wedge e_{2}$ & 0 & 0\\
\hline
$\Gamma_{\bullet}$ & $\sqrt2e_{1}$ & $0$ & $\sqrt2e_{1}\wedge e_{2}$ & 0\\
\hline
$\Gamma_{\bar\bullet}$ & 0 & $\sqrt2$ & 0 & $\sqrt2e_{2}$\\
\hline\hline
$\Gamma_{+-}$ & 1 & $e_{1}$ & $-e_{2}$ & $-e_{1}\wedge e_{2}$\\
\hline
$\Gamma_{\bar\bullet\bullet}$ & 1 & $-e_{1}$ & $e_{2}$ & $-e_{1}\wedge e_{2}$\\
\hline
$\Gamma_{+\bullet}$ & 0 & 0 & $-2e_{1}$ & 0\\
\hline
$\Gamma_{+\bar\bullet}$ & 0 & 0 & 0 & 2\\
\hline
$\Gamma_{-\bullet}$ & $-2e_{1}\wedge e_{2}$ & 0 & 0 & 0\\
\hline
$\Gamma_{-\bar\bullet}$ & 0 & $2e_{2}$ & 0 & 0\\
\hline
\end{tabular}
\end{center}
\caption{The action of the Gamma matrices and the Lorentz generators
  $\Gamma_{AB}$ on the different basis elements. \label{tab:gamma}}
\end{table}

Note that $\Gamma_A = {U_A}^a\Gamma_a$, with
\begin{displaymath}
\left({U_A}^a\right) = \frac1{\sqrt2} \left(\begin{array}{cccc} 1 & 0 & 1 & 0 \\
-1 & 0 & 1 & 0 \\ 0 & 1 & 0 & -i \\ 0 & 1 & 0 & i
\end{array}\right) \in {\mathrm U}(4)\,,
\end{displaymath}
so that the new tetrad is given by $E^A = {(U^{\ast})^A}_a E^a$.


\section{BPS equations and equations of motion}
\label{BPS-eom}

We will now show that the vanishing of the supersymmetry variations,
plus Bianchi identities and Maxwell equations, imply all equations of
motion in the timelike case, and all but one in the null case.
Without hypermultiplets, the equations of motion are (here we set $8\pi G=1$)
\begin{itemize}
\item Einstein
\begin{eqnarray}
&& 0=E_{\mu\nu}:=\frac 12 R_{\mu\nu} +(\text{Im}\,{\cal N})_{IJ} F^{+I}_{\rho \mu}{F^{-J\rho}}_{\nu} -g_{\alpha \bar \beta}
\DD_\mu z^\alpha \DD_\nu \bar z^{\bar\beta} -\frac 12 g_{\mu\nu} V;
\end{eqnarray}
\item Maxwell\footnote{We used the Bianchi identities to put the equations in this form.}
\begin{eqnarray}\label{Maxwell}
&& 0=M^\nu_I :=-2\nabla_\mu ((\text{Im}\,{\cal N})_{IJ} F^{-J\mu\nu})+i \partial_\mu {\cal N}_{IJ} \tilde F^{J\mu\nu}
-g g_{\alpha \bar \beta} k^\alpha_I \DD^\nu \bar z^{\bar\beta} \cr
&& \qquad\qquad\quad -g g_{\alpha\bar\beta} k^{\bar\beta}_I \DD^\nu z^\alpha -\frac{g^2}{4e}
C_{J,IK} \epsilon^{\nu\mu\rho\sigma} A_\mu^J F^K_{\rho\sigma};
\end{eqnarray}
\item Scalars\\
\begin{eqnarray}
&& 0=G^\alpha :=\tilde \nabla_\mu \DD^\mu z^\alpha -gA^{I\mu} \tilde \nabla_\mu k_I^\alpha +\frac 1{2i} F^{+I}_{\mu\nu} F^{+J\mu\nu}
g^{\alpha\bar \gamma} \partial_{\bar z^{\bar\gamma}}{\cal N}_{IJ} \cr
&& \quad -\frac 1{2i} F^{-I}_{\mu\nu} F^{-J\mu\nu} g^{\alpha\bar \gamma} \partial_{\bar z^{\bar\gamma}}{\bar{\cal N}}_{IJ}
-g^{\alpha\bar \gamma}\partial_{\bar z^{\bar\gamma}} { V}  \ ,\label{scalarsBPS}
\end{eqnarray}
where with $\tilde \nabla$ we mean the covariant derivative with respect to the metric connection on both the spacetime and the target
manifold of the scalars. Finally
\begin{eqnarray}
{ V}=g^2 e^{\cal K}[k^\alpha_I k^{\bar \beta}_J g_{\alpha\bar \beta} \bar Z^I Z^J +4(g^{\alpha\bar \beta} \DD_\alpha Z^I
\DD_{\bar \beta} \bar Z^J-3\bar Z^I Z^J) \vec P_I \cdot \vec P_J  ]\label{potential}
\end{eqnarray}
is the scalar potential.
\end{itemize}
We set
\begin{eqnarray}
&& \hat {\cal D}_\mu \epsilon^i=D_\mu(\omega) \epsilon^i -g\Gamma_\mu S^{ij} \epsilon_j +\frac 14 \Gamma^{ab}F_{ab}^{-I} \epsilon^{ij} \Gamma_\mu
(\text{Im}\,{\cal N})_{IJ} Z^J e^{{\cal K}/2} \epsilon_j\ .\label{demu}
\end{eqnarray}
where $D_\mu(\omega)$ is defined in (\ref{derivata}).
Then, the gravitini Killing equation is
\begin{equation}
\hat {\cal D}_\mu \epsilon^i=0\,,
\end{equation}
and its integrability is given by the (holonomy) condition
\begin{equation}
0=[\hat {\cal D}_\mu\ , \hat {\cal D}_\mu]\epsilon^i =\hat {\cal D}_\mu (\hat{\cal D}_\nu \epsilon ^i)- \hat{\cal D}_\nu (\hat{\cal D}_\mu \epsilon^i)\,.
\end{equation}
Denoting
\begin{eqnarray}
&& F_{\mu \nu}:= \partial_\mu A_\nu -\partial_\nu A_\mu \ ,\cr
&& \Phi^{ab} := Z^J e^{{\cal K}/2} (\text{Im}\,{\cal N})_{IJ} F^{-Iab}\ ,\cr
&& \bar \Phi^{ab} := \bar Z^J e^{{\cal K}/2} (\text{Im}\,{\cal N})_{IJ} F^{+Iab}\,,
\end{eqnarray}
and making use of \eqref{Gamma5-eps}, we find
\begin{eqnarray}
& 0&= \frac 14 R_{\mu\nu}^{\ \ \ ab} \Gamma_{ab}\epsilon^i +\frac i2 F_{\mu\nu} \epsilon^i +gF^I_{\mu\nu} P_{Ij}^{\ \ i} \epsilon^j
+g A_\nu^I \partial_\mu P_{Ij}^{\ \ i} \epsilon^j-g A_\mu^I \partial_\nu P_{Ij}^{\ \ i} \epsilon^j \cr
&& -2g^2 \Gamma_{\mu\nu} S^r \bar S^s \delta_{rs} \epsilon^i +2g (\bar S^i_{\ j} \Phi_{\mu\nu} -S^i_{\ j}\bar \Phi_{\mu\nu})
\epsilon^j +\left[ -\frac 12 \Phi_\nu^{\ b} \bar \Phi_\mu^{\ d} \Gamma_{bd}\right. \cr
&& \left.  +\frac 12 \Phi_\mu^{\ b} \bar \Phi_\nu^{\ d} \Gamma_{bd}
-\frac 12 \Phi^{ab} \bar \Phi_{\mu\nu} \Gamma_{ab} +\frac 12 \Phi_a^{\ b} \bar \Phi_\mu^{\ a} \Gamma_{b\nu}
-\frac 12 \Phi_a^{\ b} \bar \Phi_\nu^{\ a} \Gamma_{b\mu}\right] \epsilon^i -g\Gamma_\nu \partial_\mu S^{ij} \epsilon_j \cr
&& +g\Gamma_\mu \partial_\nu S^{ij} \epsilon_j +\frac 14 \Gamma^{ab} \epsilon^{ij} (\nabla_\mu \Phi_{ab} \Gamma_\nu
-\nabla_\nu \Phi_{ab} \Gamma_\mu)\epsilon_j-ig A_\mu \Gamma_\nu S^{ij} \epsilon_j \cr
&& +ig A_\nu \Gamma_\mu S^{ij} \epsilon_j
+\frac 14 A_\mu \Gamma^{ab} \Phi_{ab} \Gamma_\nu \epsilon^{ij}\epsilon_j -
\frac 14 A_\nu \Gamma^{ab} \Phi_{ab} \Gamma_\mu \epsilon^{ij}\epsilon_j\,.
\end{eqnarray}
Let us now contract this equation with $\Gamma^\mu$. This leads to
\begin{align}
0= &\frac 12 R_{\nu b}\Gamma^b \epsilon^i +\frac i2 \Gamma^\mu F_{\mu\nu} \epsilon^i +g\Gamma^\mu F_{\mu\nu}^I P_{Ij}^{\ \ i} \epsilon^j
+gA_\nu^I \Gamma^\mu \partial_\mu P_{Ij}^{\ \ i} \epsilon^j \cr
& -gA_\mu^I \Gamma^\mu \partial_\nu P_{Ij}^{\ \ i} \epsilon^j -6g^2\Gamma_\nu S^r S^s \delta_{rs} \epsilon^i
+2g\Gamma^\mu (\bar S^i_{\ j} \Phi_{\mu\nu}-S^i_{\ j} \bar \Phi_{\mu\nu})\epsilon^j\cr
& -2\bar \Phi_a^{\ \ b} {\Phi^a}_{\nu} \Gamma_b \epsilon^i
-g{\Gamma^\mu}_{\nu} \partial_\mu S^{ij} \epsilon_j +
3g\partial_\nu S^{ij} \epsilon_j
-igA_\mu \Gamma^\mu S^{ij}\epsilon_j \cr
& +3ig A_\nu S^{ij}\epsilon_j +(\nabla_\mu \Phi^{\mu c}+A_\mu \Phi^{\mu c})
\epsilon^{ij} (\Gamma_{c\nu}+e_{c\nu})\epsilon_j\,, \label{eccoqua}
\end{align}
where we used
\begin{eqnarray}
&& F^{I+ab} F^{J-}_{ab}=0\ ,\\
&& F^{I+}_{a[b} F^{J-a}_{\ \ \ \ c]}=0\ .
\end{eqnarray}
At this point we need to make contact with the equations of motion. To do this, let us first take the gaugini Killing equation
(multiplied with $\Gamma^{\lambda}$)
\begin{align}
0= &-2e^{{\cal K}/2} g^{\alpha \bar \beta} {\cal D}_{\bar \beta} \bar Z^I (\text{Im}\,{\cal N})_{IJ} F^{-I\lambda\mu}\Gamma_\mu \epsilon_{ij}
\epsilon^j +\Gamma^\lambda \Gamma^\mu {\cal D}_\mu z^\alpha \epsilon_i \cr
&+g\Gamma^\lambda e^{{\cal K}/2} [\epsilon_{ij}k^\alpha_I \bar Z^I-2P_{Iij}\bar {\cal D}_{\bar \beta} \bar Z^I g^{\alpha\bar \beta}]\epsilon^j\,,
\end{align}
and contract it with $e^{{\cal K}/2} {\cal D}_{\alpha} Z^L
(\text{Im}\,{\cal N})_{KL} F^{+K}_{\lambda\mu} \epsilon^{il}$.
This yields
\begin{align}
0=& -2e^{{\cal K}} g^{\alpha \bar \beta} {\cal D}_{\bar \beta} \bar Z^I {\cal D}_{\alpha} Z^L  (\text{Im}\,{\cal N})_{IJ}
(\text{Im}\,{\cal N})_{KL} F^{-I\lambda\mu} F^{+K}_{\lambda\mu} \Gamma_\mu  \epsilon^l\cr
&+e^{{\cal K}/2} {\cal D}_{\alpha} Z^L (\text{Im}\,{\cal N})_{KL} F^{+K}_{\lambda\mu}  \Gamma^\lambda \Gamma^\mu {\cal D}_\mu z^\alpha \epsilon^{il}
\epsilon_i \cr
&+g\Gamma^\lambda e^{{\cal K}} {\cal D}_{\alpha} Z^L (\text{Im}\,{\cal N})_{KL} F^{+K}_{\lambda\mu} \epsilon^{il}
[\epsilon_{ij}k^\alpha_I \bar Z^I-2P_{Iij}\bar {\cal D}_{\bar \beta} \bar Z^I g^{\alpha\bar \beta}]\epsilon^j\,. \label{questa}
\end{align}
Now add this to eq.~(\ref{eccoqua}).
Using the relation
\begin{eqnarray}
g^{\alpha\bar\beta} e^{\cal K} {\cal D}_\alpha Z^I {\cal D}_{\bar\beta} \bar Z^J =-\frac 12 (\text{Im}\,{\cal N})^{-1|IJ}
-e^{\cal K} \bar Z^I Z^J\,, \label{inverse}
\end{eqnarray}
we see that the first term of (\ref{questa}) sums up with the term $-2\bar \Phi_a^{\ \ b} \Phi^a_{\ \ \nu} \Gamma_b \epsilon^i$
of (\ref{eccoqua}) to give
\begin{eqnarray*}
(\text{Im}\,{\cal N})_{IJ} F^{+I}_{\rho \mu} F^{-J\rho}_{\ \ \ \nu}\Gamma^\mu\ .
\end{eqnarray*}
Some other useful relations are ($X^I=e^{{\cal K}/2}Z^I$)
\begin{eqnarray}
&& P_I^0=-e^{\cal K} C_{I,JK} Z^K \bar Z^J\ ,\\
&& X^J k^\alpha_J {\cal D}_\alpha X^I +iP^0_J X^J X^I =0\ ,
\end{eqnarray}
and (\ref{defN}), from which one also obtains the important identity
\begin{eqnarray}
X^J \partial_\mu {\cal N}_{IJ}=-2i{\cal D}_{\alpha}X^J{\text{Im}}\,{\cal N}_{IJ}\,
\partial_\mu z^\alpha\,.
\end{eqnarray}
After summing up (\ref{eccoqua}) and (\ref{questa}) and using the
above relations we finally find\footnote{This calculation involves a
rather mastodontic amount of algebraic manipulations, as well as the
use of some further identities of special K\"ahler geometry that can
be found in \cite{Vambroes}.}
\begin{eqnarray}
E_\nu^{\  b}\Gamma_b \epsilon^i-\frac 12 X^I M_I^\mu \epsilon^{ij} \Gamma_\mu \Gamma_\nu \epsilon_j=0\,.
\end{eqnarray}
Imposing the Maxwell equations one remains with the condition
\begin{eqnarray}
E_\nu^{\  b}\Gamma_b \epsilon^i=0\,.\label{einstein}
\end{eqnarray}
At this point one can proceed in a standard way (see for example \cite{Gauntlett:2002nw}).
If the Killing spinor is timelike, then (\ref{einstein})
implies that the Einstein equations are identically satisfied. In the other case, if the Killing spinor is null, thus selecting a null direction
``$+$'', then the equation $E_{++}=0$ must be imposed.\\
In a similar way we can handle the gaugini equations:
\begin{eqnarray}\label{scalarsgaugini}
&& 0=\delta \lambda^\alpha_i =-\frac 12 e^{{\cal K}/2} g^{\alpha \bar\gamma} \DD_{\bar\gamma} \bar Z^I (\text{Im}\,{\cal N})_{IJ} F^{-J}_{\lambda\rho}\Gamma^{\lambda\rho}
\epsilon_{ij} \epsilon^j +\Gamma^\mu \DD_\mu z^\alpha \epsilon_i +g{ N}^\alpha_{ij} \epsilon^j\ .
\end{eqnarray}
In this case the story is much longer and can be summarized as follows.
Let us first apply the operator $\Gamma^\mu{ D}_\mu (\omega)$ (see (\ref{derivata}))  to (\ref{scalarsgaugini}), contracted with $g_{\bar \beta \alpha}$.
Using (\ref{demu}) we get
\begin{align}
0=&-\frac 12 \Gamma^\mu \partial_\mu [e^{{\cal K}/2}\DD_{\bar \beta} \bar Z^I  (\text{Im}\,{\cal N})_{IJ} F^{-J}_{\lambda\rho}\Gamma^{\lambda\rho} ]
\epsilon_{ij}\epsilon^j \cr
&-\frac 18 \Gamma^\mu \omega_\mu^{ab} \Gamma_{ab}
e^{{\cal K}/2}\DD_{\bar \beta} \bar Z^I  (\text{Im}\,{\cal N})_{IJ} F^{-J}_{\lambda\rho}\Gamma^{\lambda\rho}
\epsilon_{ij}\epsilon^j\cr
&+\frac 18 \Gamma^\mu e^{{\cal K}/2}\DD_{\bar \beta} \bar Z^I  (\text{Im}\,{\cal N})_{IJ} F^{-J}_{\lambda\rho}\Gamma^{\lambda\rho}
\omega_\mu^{ab} \Gamma_{ab} \epsilon_{ij}\epsilon^j\cr
&-\frac 12 e^{\cal K} F^{-L}_{ab} F^{-Jab} \DD_{\bar \beta} \bar Z^I (\text{Im}\,{\cal N})_{IJ} (\text{Im}\,{\cal N})_{LM} Z^M \epsilon_i\cr
&+\nabla_\mu (g_{\bar \beta \alpha} \DD^{\mu} z^\alpha) \epsilon_i +2 g g_{\bar \beta \alpha} \Gamma^\mu \DD_\mu z^\alpha S_{ij}
\epsilon^j\cr
&-\frac 12 g_{\bar \beta \alpha} \Gamma^{ab} F_{ab}^{+I} (\text{Im}\,{\cal N})_{IJ}  \bar Z^J e^{{\cal K}/2}
\Gamma^\mu \DD_\mu z^\alpha \epsilon_{ij} \epsilon^j\cr
&+g \Gamma^\mu \partial_\mu ({\cal N}_{\bar \beta ij}) \epsilon^j+4 g^2 {\cal N}_{\bar \beta ij} S^{jl} \epsilon_l \ .
\label{huge}
\end{align}
At this point there are many possible manipulations which lead to the desired result. However,
the most complicated task is to recognize the derivatives of the scalar potential $V$.
To simplify such an effort, it is convenient
to express the term $\nabla_\mu (g_{\bar \beta \alpha} \DD^{\mu} z^\alpha)$ in terms of $G^\alpha$ by means of (\ref{scalarsBPS}).\\
A faster way is to work out the first term of (\ref{huge}) as follows:
\begin{eqnarray}\label{trick}
&& -\frac 12 \Gamma^\mu \partial_\mu [e^{{\cal K}/2}\DD_{\bar \beta} \bar Z^I  (\text{Im}\,{\cal N})_{IJ} F^{-J}_{\lambda\rho}\Gamma^{\lambda\rho} ]
\epsilon_{ij}\epsilon^j \cr
&& \qquad\qquad\ =-\frac 12 \Gamma^\mu \partial_\mu [e^{{\cal K}/2}\DD_{\bar \beta} \bar Z^I  (\text{Im}\,{\cal N})_{IJ}]
F^{-J}_{\lambda\rho}\Gamma^{\lambda\rho}  \epsilon_{ij}\epsilon^j \cr
&& \qquad\qquad\ \ \ \ -e^{{\cal K}/2}\DD_{\bar \beta} \bar Z^I  (\text{Im}\,{\cal N})_{IJ} \nabla_\mu F^{-J\mu \rho} \Gamma_\rho \epsilon_{ij}\epsilon^j \ ,
\end{eqnarray}
where we used the relation
\begin{eqnarray}
\Gamma_{abc}=-i\Gamma_5 \epsilon_{abcd} \Gamma^d\ ,
\end{eqnarray}
and the Bianchi identities. Then, we can use (\ref{Maxwell}) to rewrite the last term in (\ref{trick}) in terms of $M^\mu_I$, so that (\ref{huge})
takes the form
\begin{eqnarray}
g_{\bar \beta \alpha} G^\alpha \epsilon_i +\frac 12 e^{{\cal K}/2} \DD_{\bar \beta} \bar Z^I M^\nu_I \epsilon_{ij} \Gamma_\nu \epsilon^j
+\ldots =0\ .
\end{eqnarray}
Next, all the remaining manipulations are very similar to the
gravitino case, and have the aim to show that the terms indicated by dots vanish identically,
so that we will not report the details here. We only
mention that sometimes we found it convenient to use $X^I =e^{{\cal K}/2} Z^I$ in place of $Z^I$ to simplify many expressions.
Also, the Killing equations for $k^\alpha_I$ (and its conjugate) are often useful in taking account of the Christoffel symbols for
the covariant derivative on the scalar target manifold. Both (\ref{scalarsgaugini}) and its charge conjugate must be used to eliminate
many terms.\\
As we have anticipated, the final result is that (\ref{huge}) reduces to
\eqn
g_{\bar \beta \alpha} G^\alpha \epsilon_i +\frac 12 e^{{\cal K}/2} \DD_{\bar \beta} \bar Z^I M^\nu_I \epsilon_{ij} \Gamma_\nu \epsilon^j=0\ .
\feqn
Thus, if the Maxwell equations hold, the scalar fields satisfy their
equations of motion as well. Note that the results of this appendix
could also be obtained by the Killing spinor identity
approach \cite{Kallosh:1993wx,Bellorin:2005hy}.

\section{Holonomy of the base manifold}
\label{hol-base}

In order to gain a deeper geometrical understanding of the three-dimensional base space
with dreibein $V^x$, some considerations concerning its holonomy are in order.
First of all, note that in minimal ungauged ${\cal N}=2$, $D=4$ supergravity, the base is
flat \cite{Tod:1983pm} and thus has trivial holonomy. This is still true if one couples the
theory to vector multiplets \cite{Meessen:2006tu}. In five-dimensional minimal ungauged
supergravity, the base manifold is hyper-K\"ahler \cite{Gauntlett:2002nw}, whereas in the gauged
case it is K\"ahler \cite{Gauntlett:2003fk}. Thus, the general pattern in the timelike case is to
have a fibration over a base with reduced holonomy. One might therefore ask whether
our three-dimensional manifold with metric
\begin{equation}
ds_3^2=dz^2+e^{2\Phi}dw d\bar w\,, \label{3d-metric}
\end{equation}
appearing in \eqref{final-metric}, has reduced holonomy. Eqns.~\eqref{B} and
\eqref{S-FI} suggest that the Christoffel connection ${\cal A}+B$
(cf.~\eqref{dVxclosed}) has full holonomy
SO$(3)$. In fact, the only nontrivial subgroup of SO$(3)$ is U$(1)$, and integrating
the first Cartan structure equation for a Christoffel connection taking values
in u$(1)$, one finds the metric \eqref{3d-metric} with $\partial_z\Phi=0$,
which in general will not be the case.
From \eqref{red-hol}, however, it is evident that the connection ${\cal A}$ (which
has nonvanishing torsion, cf.~\eqref{dVx}), takes values in u$(1)$ $\subseteq$ so$(3)$.
The same holds for the corresponding curvature.
We can thus interpret the base space as a manifold of reduced holonomy
U$(1)$ $\subseteq$ SO$(3)$ with nonzero torsion. Reduced holonomy is equivalent to
the existence of parallel tensors, the simplest example being the reduction of
GL$(D,\bR)$ to SO$(D)$ if the metric is covariantly constant, $\nabla g=0$.
In our case, the corresponding parallel tensor is just the vector $\partial_z$:
One easily checks that $\nabla\partial_z=0$, where $\nabla$ denotes the covariant
derivative associated to ${\cal A}$.

\end{document}